\documentclass[preprint,nofootinbib,aps,superscriptaddress,eqsecnum]{revtex4-1} 
 \pdfoutput=1
 \usepackage{graphicx}
 \usepackage{amsmath}
\usepackage{amsfonts}
\usepackage{amssymb}
\usepackage{hyperref}
\usepackage{caption}
\usepackage{subcaption}
\def\bea{\begin{eqnarray}}
\def\eea{\end{eqnarray}}
 \def\be{\begin{equation}}
\def\ee{\end{equation}}
\def\nn{\nonumber}
\def\mpl{M_{\rm Pl}}

\newcommand{\MS}{\overline{\mbox{\sc ms}}}

\begin{document}

\title{Constraining minimal Type-III Seesaw Model from naturalness, Lepton Flavor Violation and Electroweak Vacuum stability}

\author{Srubabati Goswami}
\email[Email Address: ]{sruba@prl.res.in}
\affiliation{Theoretical Physics Division, 
Physical Research Laboratory, Ahmedabad - 380009, India}

 \author{Vishnudath K. N.}
\email[Email Address: ]{vishnudath@prl.res.in}
\affiliation{Theoretical Physics Division, 
Physical Research Laboratory, Ahmedabad - 380009, India}
\affiliation{Discipline of Physics, Indian Institute of Technology, Gandhinagar - 382355, India}

\author{Najimuddin Khan}
\email[Email Address: ]{khanphysics.123@gmail.com}
\affiliation{Centre for High Energy Physics, Indian Institute of Science,
C. V. Raman Avenue, Bangalore 560012, India}

\begin{abstract}
We study the minimal type-III seesaw model in which we extend the SM by adding two $SU(2)_L$ triplet fermions with zero hypercharge to explain the origin of the non-zero neutrino masses. We show that the naturalness conditions and the limits from lepton flavor violating decays provide very stringent bounds on the model parameters along with the constraints from the stability/metastability of the electroweak vacuum. We perform a detailed analysis of the model parameter space including all the constraints for both normal as well as inverted hierarchies of the light neutrino masses. We find that most of the region that are allowed by lepton flavor violating decays and naturalness fall in the stable/metastable region depending on the values of the standard model parameters.

\end{abstract}
 
\maketitle

\section{Introduction}
 
The discovery of the Higgs boson ~\cite{Chatrchyan:2012xdj,Aad:2012tfa} at the Large Hadron Collider (LHC) has confirmed the mode of generation of the masses of the fundamental particles via the mechanism of electroweak (EW) symmetry breaking and has put the Standard Model (SM) on a solid foundation. However, despite its success in explaining most of the experimental data, the SM can not address certain issues. One of the most important experimental observation that necessitates the extension of the SM is the phenomenon of neutrino oscillation. The solar, atmospheric, reactor and accelerator neutrino oscillation experiments have shown that the three  neutrino flavors mix among themselves and they have very small but non-zero masses, unlike as predicted in the SM.

Among the various beyond standard model scenarios that are proposed in the literature to explain the small neutrino masses, the most popular one is the seesaw mechanism. This is based on the assumption that the lepton number is violated at a very high energy scale by some heavier particles. The tree level exchange of these heavy particles generates the lepton number violating dimension-5 Weinberg operator $\kappa LLHH $\cite{weinberg}. This gives rise to small neutrino masses once the EW symmetry is broken. Here, $L$ and $H$ are the lepton and the Higgs doublets respectively and $\kappa$ is a proportionality constant with negative mass dimension and is inversely proportional to the energy scale at which the new physics enters. Depending on the nature of the heavy particles added for the ultraviolet completion, one can have three types of seesaw mechanism. If the seesaw is generated by adding extra neutral fermionic singlets, it is called a type-I seesaw mechanism  \cite{Minkowski:1977sc,seesaw1,seesaw2,Mohapatra:1979ia}. Similarly, type-II seesaw mechanism is generated by adding a triplet scalar \cite{Schechter:1980gr,Schechter:1981cv,Lazarides:1980nt,Mohapatra:1980yp} to the SM whereas the addition of fermionic triplets gives rise to the type-III seesaw mechanism \cite{Foot:1988aq}. It is known that in order to get a neutrino mass of the sub-eV scale, one has to take the new particles to be extremely heavy or else take the new couplings to be extremely small. This  spoils the testability of the theory. However, there are various TeV scale extensions of the canonical scenarios proposed in the literature \cite{Mohapatra:1986bd,Gu:2010xc,Zhang:2009ac,Hirsch:2009mx,Gavela:2009cd} which can be probed at the collider experiments (For recent reviews see for 
instance \cite{Boucenna:2014zba,Deppisch:2015qwa}). In the case of type-I and type-III seesaw models, one can also have large Yukawa couplings and new fermions of masses in the TeV scale by choosing some particular textures of the neutrino Yukawa coupling matrix  \cite{Adhikari:2010yt,Kersten:2007vk,Pilaftsis:1991ug}.

An important aspect to be considered while studying the seesaw models is the issue of naturalness. It is well known that  the Higgs mass gets large corrections from the higher order loop diagrams due to its self-interaction as well as the couplings with gauge bosons and fermions. The theory is perceived unnatural if a severe fine-tuning
between the quadratic radiative corrections and the bare mass is needed to bring down Higgs mass to the observed scale.
It is well known that although the dimensional regularization can throw away the quadratic divergences, the presence of other dangerous logarithmic and finite contributions can cause similar naturalness problem. In the case of seesaw models in which the new particles couple to the SM Higgs, this naturalness problem is enhanced ~\cite{Vissani:1997ys,Casas:2004gh,Abada:2007ux,Farina:2013mla,Clarke:2015gwa,
Fabbrichesi:2015zna,Clarke:2015hta,Chabab:2015nel,Haba:2016zbu,Bambhaniya:2016rbb,Dev:2017ouk}. Demanding the correction to the Higss mass to be of the order of TeV can bring down the seesaw scale.

 Another aspect of low scale seesaw models which has received attention lately is the implications of such scenarios for the stability of the EW vacuum. It is to be noted that the observed value of the Higgs mass of $125.7\pm0.3$ GeV is quite intriguing from the viewpoint of the EW vacuum stability. The measured values of the SM parameters, especially the top mass $M_t$ and strong coupling constant $\alpha_s$ suggest that an extra deeper minima resides near the Planck scale, threatening the stability of the present EW vacuum \cite{Alekhin:2012py,Buttazzo:2013uya}, since this may tunnel into that true (deeper) vacuum. The decay probability has been calculated using the state of the art NNLO corrections and it suggests that the
present EW vacuum is metastable at $~3\sigma$ which means that the decay time is greater than the age of the universe. It is well known that the scalar couplings pull the vacuum towards stability whereas the Yukawa couplings push it towards instability. Thus, in the case of seesaw models, the Yukawa couplings as well as the masses of the new fermions will also get bounded by the constraints from the stability/metastability of the EW vacuum \cite{Rodejohann2012,Chakrabortty:2012np,Khan:2012zw,Datta:2013mta,Chakrabortty:2013zja,Kobakhidze:2013pya,Ng:2015eia,
Rose:2015fua,Bambhaniya:2014hla,Lindner:2015qva,Das:2015nwk,Das:2016zue,Bhattacharya:2017fid,Garg:2017iva}. In particular, in reference \cite{He:2012ub}, the authors have discussed the implications of vacuum stability and gauge-Higgs unification in the context of the type-III seesaw model and reference \cite{Lindner:2015qva} has discussed the EW vacuum metastability in the context of type-I as well as type-III seesaw models. In reference \cite{Bambhaniya:2016rbb}, the authors have studied the implications of naturalness and vacuum stability in a minimal type-I seesaw model. Similarly, the naturalness and vacuum stability in the case of the type-II seesaw model have been studied in reference \cite{Haba:2016zbu}.

 In this paper, we study the consequences of naturalness in the minimal type-III seesaw model, in which we extend the SM by adding two $SU(2)_L$ triplet fermions with zero hypercharge to explain the origin of the non-zero neutrino masses and mixing. To give mass to all the three light active neutrinos, one needs three triplet fermions. Hence, in the minimal type-III seesaw model, the lightest active neutrino will be massless. We use the Casas-Ibarra (CI) parametrization for the neutrino Yukawa coupling matrix \cite{Casas:2001sr,Ibarra:2003up} and by choosing the two triplets to be degenerate, we have only three independent real parameters, namely the mass of the triplet fermions and a complex angle in the CI parametrization. We study and constrain the bounds on these model parameters by demanding the theory to be natural. In addition, we also study the bounds on the model from the EW vacuum metastability as well as lepton flavor violating (LFV) decays.

The rest of the paper is organized as follows: In Sec. II, we review the minimal type-III seesaw model and the parametrization used for our studies. In Sec. III, we discuss the implications of naturalness in the minimal type-III seesaw model and in section IV, we have discussed the constraints from the LFV decays. After this, we discuss the effective Higgs potential in the presence of the extra fermion triplets and the renormalization group (RG) evolution of the different couplings, and present a detailed discussion of the results. Finally, we summarize in Sec. VI.

\section{The MInimal Type-III Seesaw Model}

We extend the standard model with two fermionic triplets $\Sigma_{R_{i}}$, $ i=1,2$ having zero hypercharge, which can be represented as,
\be  \Sigma_R \,=\, \begin{bmatrix}
    \Sigma_R^0/\sqrt{2} & \Sigma_R^+ \\
    \Sigma_R^- & -\Sigma_R^0/\sqrt{2}
\end{bmatrix} \, \equiv \,\frac{\Sigma_R^i \sigma^i}{\sqrt{2}},
\ee
where $\Sigma_{R}^{\pm} \, = \, (\Sigma_{R}^1 \mp i \Sigma_{R}^2)/\sqrt{2}$.
The parts of the Lagrangian that are relevant to neutrino mass generation are,
\be
- {\cal L}_\Sigma = \tilde{\Phi}^\dag \overline{\Sigma}_R\sqrt{2}Y_\Sigma L + \frac{1}{2} ~\textrm{Tr}~[\overline{\Sigma}_R M \Sigma_R^c]   \,\,\, + \,\,\, \textrm{h.c.},
\ee
where the generation indices have been suppressed. In the above equation, $L=(\nu_l \, \, \, l^-)^T$ is the lepton doublet and $\tilde{\Phi} = i\sigma_2 \Phi^*$ ($\sigma_2$ is the second Pauli matrix). For simplicity, we consider the scenario in which the Majorana mass matrix $M$ is proportional to the identity matrix so that the heavy fermions have degenerate masses, denoted by $M_\Sigma$. Once the Higgs field $\Phi$ acquires a vacuum expectation value ($vev$), the neutral fermion mass matrix can be written as, 
\be M_\nu = \begin{pmatrix}
  0 & M_D^T  \\
 M_D & M
\end{pmatrix} .
\ee
Here, $M_D = Y_\Sigma v /\sqrt{2}$, where $v = 246$ GeV is the $vev$ of the SM Higgs. The given mass matrix can be diagonalized by a unitary matrix $U_0$ as,
\be
U_0^T M_\nu U_0 = M_\nu^\textrm{diag} = \textrm{diag}(m_1,~m_2~,m_3~,M_\Sigma,~M_\Sigma),
\ee
where $M$ is the mass of the heavy triplet fermions. Note that the lightest neutrino is massless in this scenario. We can write the matrix $U_0$ as \cite{Grimus:2000vj} , 
\be
U_0 = 	W U_\nu \simeq  \begin{pmatrix}
  (1-\frac{1}{2}\epsilon) U & M_D^\dag (M^{-1})^* U_R  \\
 -M^{-1} M_D U & (1-\frac{1}{2}\epsilon') U_R
\end{pmatrix}  \equiv  \begin{pmatrix}
  U_L & T  \\
 S & U_H
\end{pmatrix}.
\ee
Here, $W$ brings the full $5 \times 5$ mass matrix to the block diagonal form and $U$ and $U_R$ diagonalizes the light and heavy neutrino mass matrices respectively. In our case, $U_R$ is $2\times 2$ identity matrix. $U_L$ is the Pontecorvo-Maki-Nakagava-Sakata (PMNS) mixing matrix with a
small non-unitary correction. The non-unitarity is characterized by $\epsilon$ and $\epsilon'$  and are given by,
\be
\epsilon = TT^\dag = M_D^\dag (M^{-1})^* M^{-1} M_D \, \, , \, \,\,\, \, \epsilon' = SS^\dag = M^{-1} M_D M_D^\dag (M^{-1})^*\,.
\ee 
In the limit $M >> M_D$, the light neutrino mass matrix can be written as,
\be
m_{\textrm{light}} = -M_D^T M^{-1} M_D.
\ee
We use the Casas-Ibarra parametrization \cite{Casas:2001sr,Ibarra:2003up} for the Yukawa coupling matrix $Y_\Sigma$, such that the constraints on the light neutrino mixing angles as well as the mass squared differences as predicted from the oscillation data are automatically satisfied. In this parametrization,
\be
Y_\Sigma = \frac{\sqrt{2}}{v}\sqrt{D_\Sigma} R \sqrt{D_\nu} U^\dag, \label{ciy}
\ee
where $D_\Sigma = \textrm{diag}(M_\Sigma, M_\Sigma) $, $D_\nu = \textrm{diag}(m_1, m_2, m_3)$, and $R$ is an arbitrary complex $2 \times 3$ orthogonal matrix which parametrizes the information that is lost in the decoupling of the triplet fermions. The light neutrino masses for the normal and inverted hierarchies are given by,
\be
 m_1=0 \,,\,\, m_2= \sqrt{\Delta m_{sol}^2}\,, \,\, m_3= \sqrt{\Delta m_{atm}^2} \,\,\,\,\,\, (\textrm{NH}) \nonumber \ee
\be
m_1=\sqrt{\Delta m_{atm}^2} \,,\,\, m_2= \sqrt{\Delta m_{sol}^2+\Delta m_{atm}^2}\,, \,\, m_3= 0 \,\,\,\,\,\, (\textrm{IH}). \label{NHIH}
\ee
We use the following parametrization of the PMNS matrix $U$ :,
\be
U = \begin{pmatrix}
  c_{12}c_{13} & s_{12}c_{13} & s_{13} e^{-i\delta} \\
 -c_{23}s_{12}-s_{23}s_{13}c_{12}e^{i\delta} & c_{23}c_{12}-s_{23}s_{13}s_{12}e^{i\delta} & s_{23}c_{13} \\
 s_{23}s_{12}-c_{23}s_{13}c_{12}e^{i\delta} & -s_{23}c_{12}-c_{23}s_{13}s_{12}e^{i\delta} & c_{23}c_{13}
\end{pmatrix} P,
\ee
where $ c_{ij} = cos\theta_{ij} \,\, , \,\, s_{ij} = sin\theta_{ij}$ and the phase matrix 
$P = \textrm{diag}\, (e^{-i\alpha}, \, e^{+i\alpha },\,1)$  contains the Majorana phases.

\begin{table}[ht]
 $$
 \begin{array}{|c|c|c|}
 \hline {\mbox {Parameter} }& NH & IH \\
 \hline
  \Delta m^2_{21}/10^{-5} eV^{2}&            6.80 \rightarrow 8.02 &  6.80 \rightarrow 8.02 \\
   \Delta m^2_{3l}/10^{-3} eV^{2}&             +2.399 \rightarrow +2.593 &  -2.562 \rightarrow -2.369\\
          \sin^2\theta_{12}&            0.272\rightarrow 0.346  & 0.272\rightarrow 0.346\\
          \sin^2\theta_{23}&            0.418 \rightarrow 0.613  &  0.435 \rightarrow 0.616\\
          \sin^2\theta_{13}&            0.01981 \rightarrow 0.02436 &  0.02006 \rightarrow 0.02452\\
                
 \hline
 \end{array}
 $$
 \label{table f}\caption{\small{ The oscillation parameters in their $3\sigma$ range,  for both NH and IH as given by the global analysis
of neutrino oscillation data with three 
light active neutrinos \cite{Esteban:2016qun}. 
 }}\label{oscdata}\end{table}

In our numerical analysis, we have used the values of mass squared differences and mixing angles in the $3 \sigma$ ranges as shown in table (\ref{oscdata}) \cite{Esteban:2016qun} and varied the phases $\delta$ and $\alpha$ between $-\pi$ to $+\pi$. It has been shown in reference \cite{Ibarra:2003up} that the matrix $R$  can be parametrized as,
\be
R =
  \begin{cases}
    \begin{pmatrix}
  0 & \,\,\,\,\, \textrm{cos}~z & \, \,\,\,\,\zeta ~\textrm{sin}~z  \\
 0 & \,\,\,\,\, -\textrm{sin}~z & \, \,\,\,\,\zeta ~\textrm{cos}~z  
\end{pmatrix} \,\,\,\, (\textrm{NH}) \\
    \begin{pmatrix}
   \textrm{cos}~z & \, \,\,\,\,\zeta ~\textrm{sin}~z & \,\,\,\,\, 0 \\
-\textrm{sin}~z & \, \,\,\,\,\zeta ~\textrm{cos}~z  & \,\,\,\,\,0 
\end{pmatrix} \,\,\,\, (\textrm{IH}), \\
  \end{cases}
  \ee
where $z$ is a complex parameter and $\zeta = \pm 1$. We fix the value of $\zeta$ to be $+1$ for all our analysis and this doesn't change any of our results. Thus the only free parameters in the model are the mass of the triplet fermions, $M_\Sigma$ and the complex number, $z$. $z$ can take any value in the complex plane.

Note that in this model, the charged components of the triplet fermions mix with the SM charged leptons. This is governed by the Lagrangian \cite{Abada:2008ea},

\be L = -  \begin{pmatrix}
  \bar{l}_R   & \bar{\Psi}_R
\end{pmatrix} \begin{pmatrix}
  m_l & 0  \\
 \sqrt{2} M_D & M
\end{pmatrix} 
  \begin{pmatrix}
  {l_L}   \\
 {\Psi_L}
\end{pmatrix} \,\,\,\,\,+\,\,\,\,\,\textrm{h.c.}, \ee

where we have defined, 

\be \Psi = \Sigma_R^{+c} \,\,+ \,\,\Sigma_R^- . \ee

The charged fermion mass matrix given in the above equation can be diagonalized by a bi-unitary transformation.

 Since the additional heavy triplet fermions have the $SU(2)$ gauge interactions, they can be produced and detected in the collider experiments through the process(es) $pp \rightarrow \Sigma^+ \Sigma^- \rightarrow m\, j + n\,l + \ensuremath{{\not\mathrel{E}}_T}$ ($m, n$ are integers). The collider study of extra triplet fermions was first explored in reference \cite{Bajc:2006ia} in the context of a SU(5) GUT model. Since then, a lot of works have been done on the phenomenology of type-III seesaw model in the context of LHC \cite{Bajc:2007zf,Franceschini:2008pz,delAguila:2008cj, Li:2009mw,Bandyopadhyay:2011aa,Eboli:2011ia,vonderPahlen:2016cbw,Ruiz:2015zca}. The experimental searches performed by the CMS and the ATLAS have put lower bounds on the triplet masses.
CMS \cite{CMS:2016hmk} has set a lower limit of 430 GeV on the triplet
mass with the data from $\sqrt{s} = 13$ TeV 
run whereas depending on the various scenarios studied, the ATLAS results rule out masses in the range below $325 - 540$ GeV \cite{Aad:2015cxa}. Recently, the authors of reference \cite{Goswami:2017jqs} have studied the phenomenology of type-III seesaw model in the context of high energy $e^+e^-$ colliders.

\section{Naturalness}

One of the problems associated with the high-scale seesaw models is that the associated heavy particles give very large corrections to the Higgs mass making the theory unnatural. Here, we shall see the implications of naturalness in the context of the type-III seesaw scenario. The tree level SM Higgs potential is given by,
 \be
  V = -\mu^2(\Phi^\dag \Phi) +\lambda (\Phi^\dag \Phi)^2,
 \ee
 
 where,
 
 \be   \Phi = \frac{1}{\sqrt{2}}\begin{pmatrix}
  G^{+}\\
 v+h+iG^0
\end{pmatrix}.   \ee
 Here, the $vev$, $v = 246 ~\textrm{GeV}$ and this will give the physical Higgs particle with tree level mass as $m_h^2= 2 \lambda v^2$. For the naturalness of the Higgs mass, the heavy right handed neutrino loop corrections to the mass parameter $\mu$ should be smaller than $O(\textrm{TeV}^2)$. In the $\overline{\textrm{MS}}$ scheme, the correction is given by,
 \be
 \delta \mu^2 \approx \frac{3}{4 \pi^2} \textrm{Tr} [Y_\Sigma^\dag D_\Sigma^2 Y_\Sigma].\ee
 
 Note that we have taken the quantity $\large (\textrm{ln} \large[\frac{M_\Sigma}{\mu_R}\large] - \frac{1}{2}\large )$ to be unity (where $\mu_R$ is the renormalization scale). Now, using the parametrization in eqn.(\ref{ciy}), we get,
 \be \delta \mu^2 \approx \frac{3}{4\pi^2}\frac{2}{v^2}~\textrm{Tr}[D_\nu R^\dag D_\Sigma^3 R] ~=~\frac{3 M_\Sigma^3}{2\pi^2 v^2}\textrm{cosh}(2\textrm{Im}[z])\times \begin{cases}
    \sqrt{\Delta m_{sol}^2} + \sqrt{\Delta m_{atm}^2} \,\,\,\,\,\,\,\,\,\,\,\,\,\,\,\,\,\,\,\,\,\,\,\,\,\,\,\,\, (NH) \\
    \sqrt{\Delta m_{atm}^2} + \sqrt{\Delta m_{sol}^2+\Delta m_{atm}^2} \,\,\,\,\, (IH). \\
  \end{cases} \label{nat}
  \ee

From the above expressions, we can see that the only unknown parameters are $M_\Sigma$ and $\textrm{Im}[z]$.
 
\begin{figure}[tbh]
    \centering
    \begin{subfigure}[b]{0.49\textwidth}
        \includegraphics[width=\textwidth]{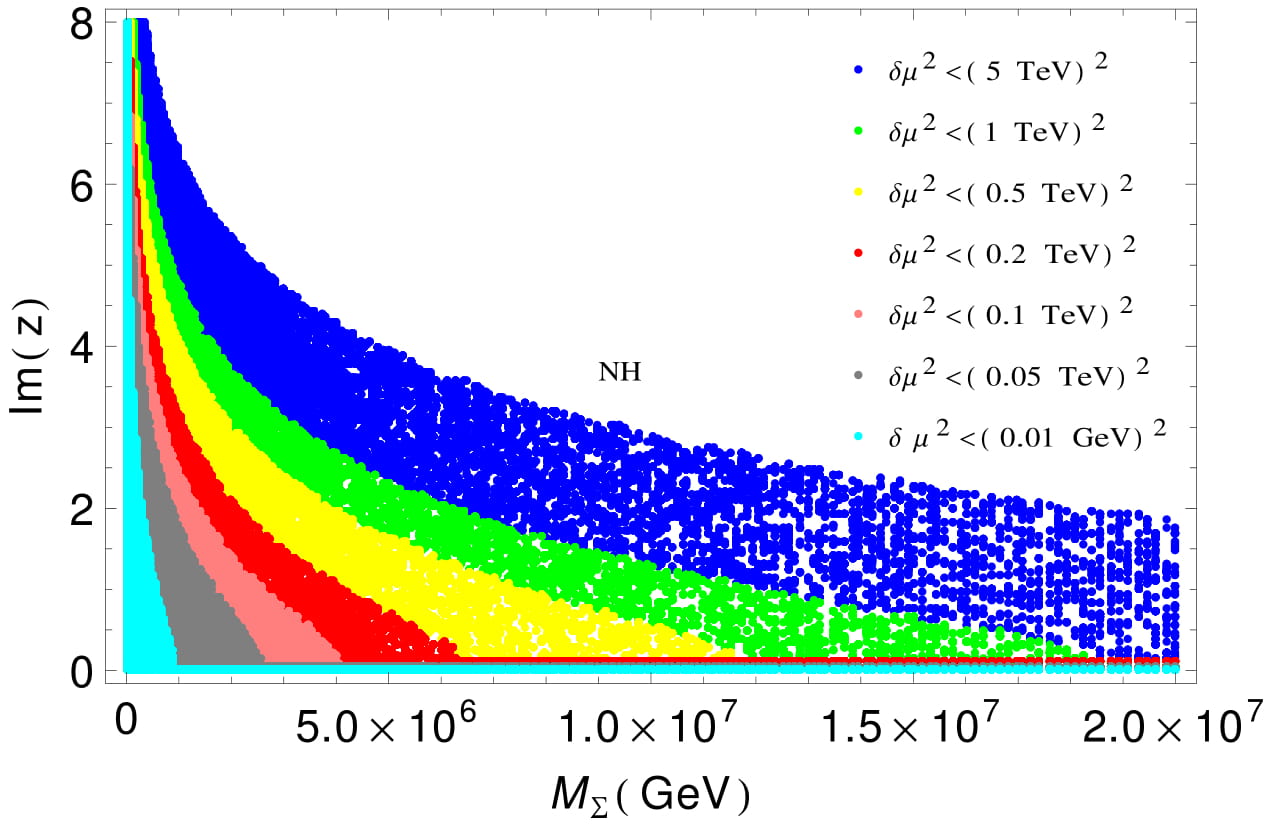}
        \caption{ \centering Naturalness contour for NH} \label{1a}
    \end{subfigure}
    ~ 
    \begin{subfigure}[b]{0.49\textwidth}
        \includegraphics[width=\textwidth]{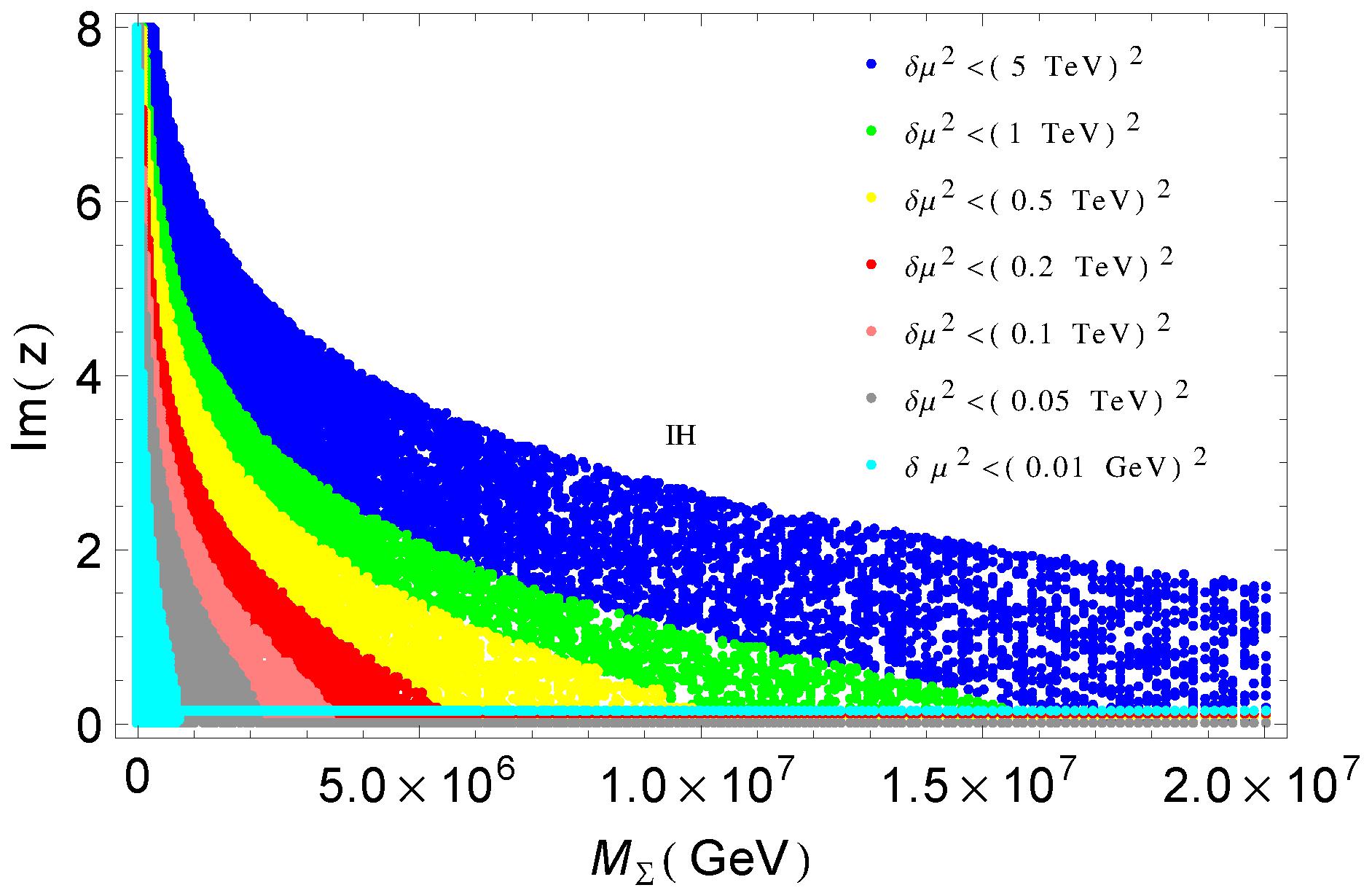}
        \caption{\centering Naturalness contours for IH} \label{1b}
            \end{subfigure}
\caption{\centering  Naturalness contours in the Im$[z]$-$M_\Sigma$ plane. The left (right) plot is for NH (IH). In the shaded rgions, $\delta\mu^2$ is less than $p\%$ of $1 \textrm{TeV}^2$ where $p=500,100,50,20,10,5,1$ (from top to bottom). The unshaded regions are disfavored by naturalness. }\label{1}
\end{figure}
In Fig.\ref{1}, we have presented the naturalness contours in the Im$[z]$-$M_\Sigma$ plane for both NH and IH. In the shaded rgions, $\delta\mu^2$ is demanded to be less than $p\%$ of $1 \textrm{TeV}^2$ where $p=500,100,50,20,10,5,1$ (from top to bottom). The unshaded regions are disfavored by naturalness. From these plots, we can see that higher the mass of the triplet, smaller the allowed values of the Im $[z]$. For instance, demanding $\delta \mu^2 < (1 ~\textrm{TeV})^2$ implies that $M_\Sigma \leq 1.84 \times 10^7 $ GeV for $\textrm{Im}[z]=0$ and $M_\Sigma \leq 3 \times 10^5 $ GeV for $\textrm{Im}[z]=6$. These bounds become even more stringent as we demand $\delta\mu^2$ to be smaller as could be seen from the plots. Also, from eqn.(\ref{nat}), we can see that the $\delta\mu^2$ values for NH and IH differ roughly by a factor of half ($\Delta m_{atm}^2 >> \Delta m_{sol}^2$). This effect can be seen from the fact that for a given value of Im$(z)$, the maximum allowed value of $M_\Sigma$ for NH is slightly higher than that for IH.

\section{Constraints from the LEPTON FLAVOUR VIOLATION}

The decay widths and the branching ratios (BR) for the various lepton flavor violating decays in the context of type-III seesaw model have been worked out in the reference \cite{Abada:2008ea}.  This model can have the decays $\mu \rightarrow e \gamma $ and $\tau \rightarrow l \gamma$ at the one loop level and $\mu \rightarrow 3e$ as well as $\tau \rightarrow 3 l$ processes in the tree level due to the charged lepton mixing. However, among all the LFV decays, the most stringent bound is the one coming from $\mu$ to e conversion  in the nuclei. The $\mu \rightarrow e$ conversion rate to the total nucleon muon capture rate ratio ($R^{\mu \rightarrow e}$) puts a bound on $\epsilon_{e \mu}$. For the ${}_{22}^{48}$Ti nuclei, we have~\cite{Dohmen:1993mp}, \be  R^{\mu \rightarrow e} < 4.3 \times 10^{-12}, \ee  and the bound from this is the most stringent among all the LFV bounds in the triplet fermion model  and is given as \cite{Abada:2008ea},
 \be \epsilon_{e \mu} < 1.7\times 10^{-7}.  \label{lfvbound}\ee

\begin{figure}[tbh]
    \centering
    \begin{subfigure}[b]{0.49\textwidth}
        \includegraphics[width=\textwidth]{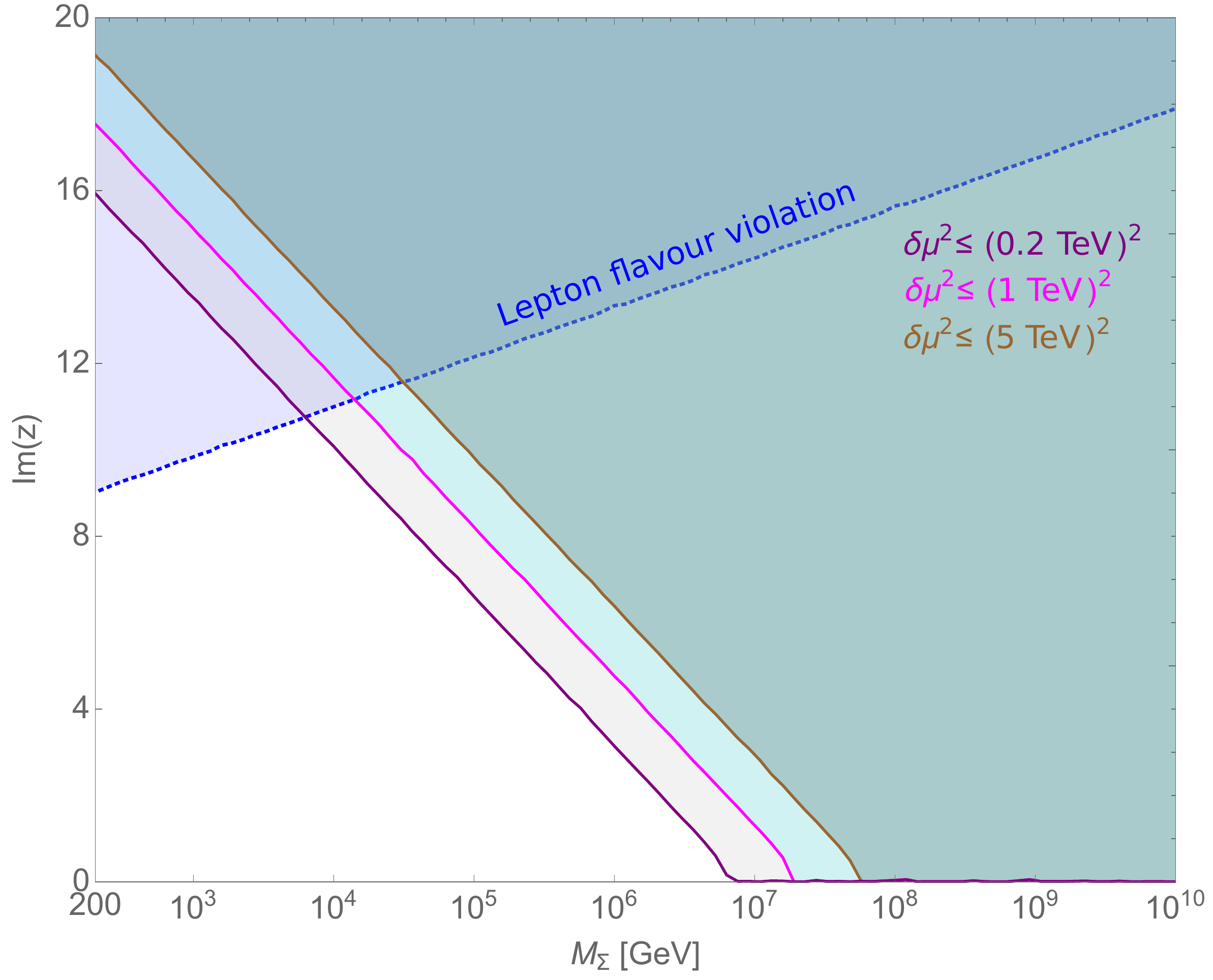}
        \caption{ \centering NH} \label{2a}
    \end{subfigure}
    ~ 
    \begin{subfigure}[b]{0.49\textwidth}
        \includegraphics[width=\textwidth]{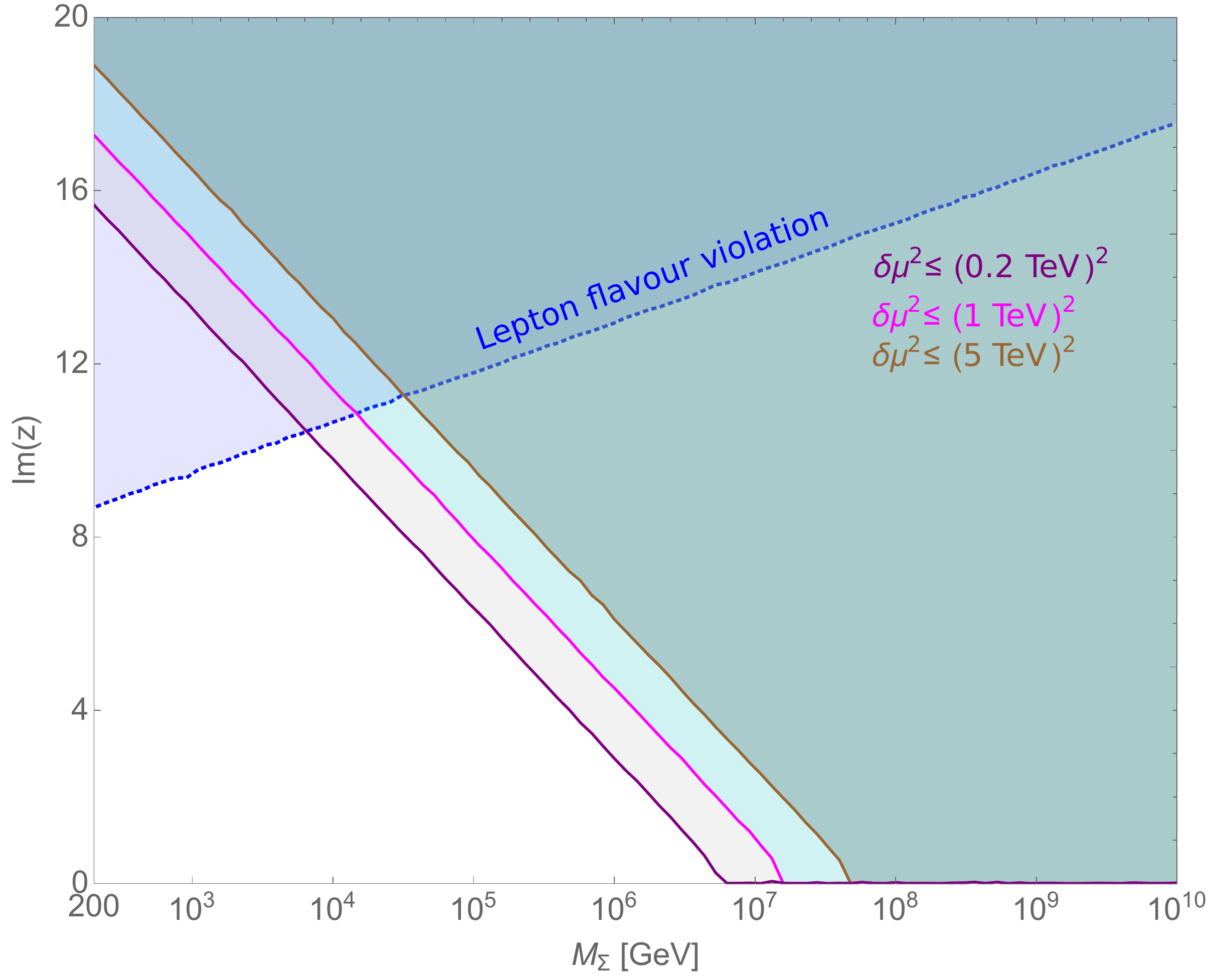}
        \caption{\centering IH} \label{2b}
            \end{subfigure}
    
     \caption{\centering  Bounds on $z$ from lepton flavour violation (blue dotted line) and naturalness (purple, magenta and brown solid lines). The left (right) plot is for NH (IH). The unshaded region is allowed by both LFV as well as naturalness bounds. }\label{3}
\end{figure}

We present the constraints on $z$ and $M_\Sigma$ from this bound in Fig.\ref{3} for both NH and IH.
The region above the blue dotted line are disallowed by the LFV bounds whereas the regions to the right of the purple, magenta and brown solid lines are disallowed by the naturalness bounds depending on the naturalness condition used. We can clearly see that the naturalness bounds restrict larger values of $M_\Sigma$ whereas the LFV bound constrains the larger values of Im$(z)$ corresponding to the smaller values of $M_\Sigma$. The unshaded region is the one that is allowed by both LFV as well as the naturalness bounds. One can notice from these plots that for both NH and IH, the maximum allowed value of Im$(z)$ is $\sim 10$ which corresponds to a triplet mass of $\sim 10^4$ GeV.
In generating these plots, we have varied the light neutrino mass squared differences and mixing angles in their $3\sigma$ ranges and the Dirac and Majorana phases are varied in the  range $0-\pi$ and we have presented the most stringent bounds.

\section{Vacuum Stability}
 In this section, we discuss how the stability of the EW vacuum is modified in the presence of the extra fermionic triplets if we assume that there is no other new physics up to the Planck scale. It is well known that if we have extra fermions, they tend to destabilize the EW vacuum. We aim to quantify this effect and obtain constraints in the context of the model outlined. In the following, we discuss the theoretical background and tools needed in the stability analysis of the EW vacuum up to the Planck scale such as the Higgs effective potential which determines the instability, metastability, stability and perturbative-unitary scales, the proper matching conditions which give the initial values of the model parameters at the electroweak (EW) scale and the RGEs delineating the running of the couplings and the other parameters from the EW scale up to the Planck scale $\mpl$.

The SM one-loop effective Higgs potential in the $\MS$ scheme and the Landau gauge can be written as
\be
V_1^{\rm SM}(h)=\sum_{i=1}^5 \frac{n_i}{64 \pi^2} M_i^4(h) \left[ \ln\frac{M_i^2(h)}{\mu^2(t)}-c_i\right] \,,
\ee

where the index $i$ is summed over all SM particles,  $M_i^2(h)= \kappa_i(t)\, h^2(t)-\kappa_i^{\prime}(t)$ and $c_{h,G,f}=3/2$, $c_{W,Z}=5/6$ ~\cite{Casas:1994qy, Altarelli:1994rb, Casas:1994us, Casas:1996aq, Quiros:1997vk}. $n_i$ is the number of degrees of freedom of the particle fields.  The values of  $n_i$, $\kappa_i$ and $\kappa_i'$ are given in the eqn.(4) in \cite{Casas:1994qy}. The above contribution comes with a positive sign for the gauge and scalar bosons, whereas it is  negative for the fermion fields. The running energy scale $\mu$ is related to a dimensionless parameter $t$ as $\mu(t)=M_Z \exp(t)$.

Following the method outlined in ~\cite{Casas:1999cd,Khan:2012zw,Garg:2017iva}, the additional contribution to the one-loop effective potential from the fermionic triplet is given as,
\be V_1^{\Sigma} (h) = - \frac{3 (M_{D}^\dag (h) M_{D}(h))_{ii}^2}{32 \pi^2} \Bigg[ \textrm{ln}\,\frac{(M_{D}^\dag (h)M_{D}(h))_{ii}}{\mu^2(t)}  -\frac{3}{2}\Bigg] -  \frac{3 (M_{D}(h) M_{D}^\dag (h))_{jj}^2}{32 \pi^2} \Bigg[ \textrm{ln}\,\frac{(M_{D}(h) M_{D}^\dag (h))_{jj}}{\mu^2(t)}  -\frac{3}{2}\Bigg] ,
 \ee
where $M_{D}(h) = \frac{Y_\Sigma}{\sqrt{2}}\,h$ and $i$, $j$ run over the three light neutrinos and the two triplet fermions respectively. In this analysis, we use the two-loop contributions to the effective potential for the SM particles whereas the contribution due to the extra fermion triplet is considered up to one-loop only.
For high field value $h(t) \, >> \, v$, the effective potential can be approximated as, $ 
V_{eff}^{SM+\Sigma}\, = \, \lambda_{eff}(h) \frac{h^4}{4}
$. The one- and two- loop SM expressions for $\lambda_{eff}(h)$ can be found in reference~\cite{Buttazzo:2013uya}.
The contribution due to the extra fermionic triplet is obtained as,

\be \lambda^{\Sigma}_{eff}(h) \, = \, -  \frac{3\,e^{4 \Gamma (h)}}{32 \pi^2} \Bigg( (Y_\Sigma^\dag Y_\Sigma)_{ii}^2  \Big(   \textrm{ln}\,\frac{({Y}_\Sigma^\dag {Y}_\Sigma)_{ii}}{2} \,-\frac{3}{2} \Big) +  (Y_\Sigma Y_\Sigma^\dag)_{jj}^2  \Big(   \textrm{ln}\,\frac{(Y_\Sigma Y_\Sigma^\dag)_{jj}}{2} \,-\frac{3}{2}\ \Big) \Bigg)
\label{labmbdaeffnu}
\ee
where, the factor $\Gamma(h) \, = \, \int_{M_t}^{h} \gamma(\mu)\, d\,\textrm{ln}\,\mu $
indicates the wave function renormalization of the Higgs field. Here $\gamma(\mu)$ is the anomalous dimension of the Higgs~\cite{Casas:1994qy, Altarelli:1994rb, Casas:1994us, Casas:1996aq, Quiros:1997vk}, the contribution to which from the fermion triplet at one loop is $\frac{3}{2} \mbox{Tr}\Big({Y_\Sigma  Y_{\Sigma}^{\dagger}}\Big)$. We also assume that $\mu=h$. In this choice, all the running coupling constants ensure faster convergence of the perturbation series of the potential~\cite{Ford:1992mv}.

We compute the RG evolution of all the couplings to analyse the Higgs potential up to the Planck Scale. We first calculate all the SM couplings at the top mass scale $M_t$, taking care of the threshold corrections~\cite{Sirlin:1985ux,Bezrukov:2012sa,Degrassi:2012ry,Khan:2014kba}.
We use one-loop RGEs to calculate $SU(2)$ and $U(1)$ gauge couplings $g_2(M_t)$ and $g_1(M_t)$ \footnote{Our result will not change significantly even if we use the two-loop RGEs for $g_1$ and $g_2$.}. For the $SU(3)$ gauge coupling $g_3(M_t)$, we use three-loop RGEs considering contributions from the five quarks and the effect of the sixth, i.e., the top quark has been taken using an effective field theory approach.
We also include the leading term in the four-loop RGE for $\alpha_s$. The mismatch between the top pole mass and the $\overline{MS}$ renormalized coupling has been taken care by using the threshold correction  $ y_t(M_t) \, = \, \frac{\sqrt{2}M_t}{v}\,(1\,+\,\delta_t(M_t)) $,
where $\delta_t(M_t)$ is the matching correction for $y_t$ at the top pole mass.  We use $\lambda(M_t) \, = \, \frac{M_H^2}{2 v^2} \, (1\,+\,\delta_H(M_t)) $ for the Higgs quartic coupling $\lambda$. To calculate this at the scale $M_t$,
we have included the QCD corrections up to three loops \cite{Melnikov:2000qh}, electroweak corrections up to one-loop \cite{Hempfling:1994ar,Schrempp:1996fb} and the $O(\alpha \alpha_s)$ corrections to the matching of top Yukawa and top pole mass \cite{Bezrukov:2012sa,Jegerlehner:2003py}.
We have reproduced the SM couplings at $M_t$ as in references \cite{Buttazzo:2013uya,Khan:2014kba} by using these threshold corrections.
We evolve them up to the heavy fermionic mass scale using the SM RGEs~\cite{Chetyrkin:2012rz,Zoller:2012cv,Chetyrkin:2013wya,Zoller:2013mra}. The extra contributions due to the femionic triplets are included after the threshold heavy fermionic mass scale~\cite{Chakrabortty:2008zh}.
Then we evolve all the couplings up to the Planck scale to find the position and depth of the new minima at the high scale.

It is well known that if the EW vacuum of the Higgs potential is not the global minimum, then a quantum tunneling to the true vacuum may occur. This happens because the RG running can make the quartic coupling $\lambda$ negative at a high energy scale. However, this does not pose a threat to the theory if the decay time is greater than the lifetime of the Universe $\tau_U\sim 10^{17}$ secs~\cite{Ade:2015xua} and in such a case, we say that the EW vacuum is metastable. The decay probability of the EW vacuum to the true vacuum at the present epoch has been computed using
the bounce solution of the euclidean equations of motion of the
Higgs field~\cite{Coleman:1977py, Isidori:2001bm, Buttazzo:2013uya},
\be
{\cal P}_0=0.15 \frac{\Lambda_B^4}{H^4} e^{-S(\Lambda_B)}, \,\,{\rm where}\,\,S(\Lambda_B)=\frac{8\pi^2}{3|\lambda_{eff}(\Lambda_B)|}\, .
\label{eq:probaction}
 \ee
Here, $H$ is the Hubble constant and $S(\Lambda_B)$ is the minimum action of the Higgs potential at the bounce size $R=\Lambda_B^{-1}$ which gives the dominant contribution to the tunneling probability ${\cal P}_0$. The metastable EW vacuum implies that the decay probability ${\cal P}_0 < 1$.
This can be translated into a 
bound on the Higgs effective quartic coupling $\lambda_{eff}$ which can be read as \cite{Isidori:2001bm,Khan:2014kba,Khan:2015ipa},
\be \lambda_{ eff} > \lambda_{ eff~ \rm min}(\Lambda_B) = \frac{-0.06488}{1-0.00986 \ln\left( {v}/{\Lambda_B} \right)} .\ee
 $\lambda_{eff}(\Lambda_B)< \lambda_{ eff~ \rm min}(\Lambda_B)$ corresponds to the unstable region and the EW vacuum is absolutely stable at $\lambda_{eff}(\Lambda_B) > 0$. Also, the theory violates the perturbative unitarity at $\lambda_{eff}(\Lambda_B) > \frac{4\pi}{3}$~\cite{Lee:1977eg}.

  \begin{figure}[h!]
    \begin{subfigure}[b]{0.469\textwidth}
        \includegraphics[width=\textwidth]{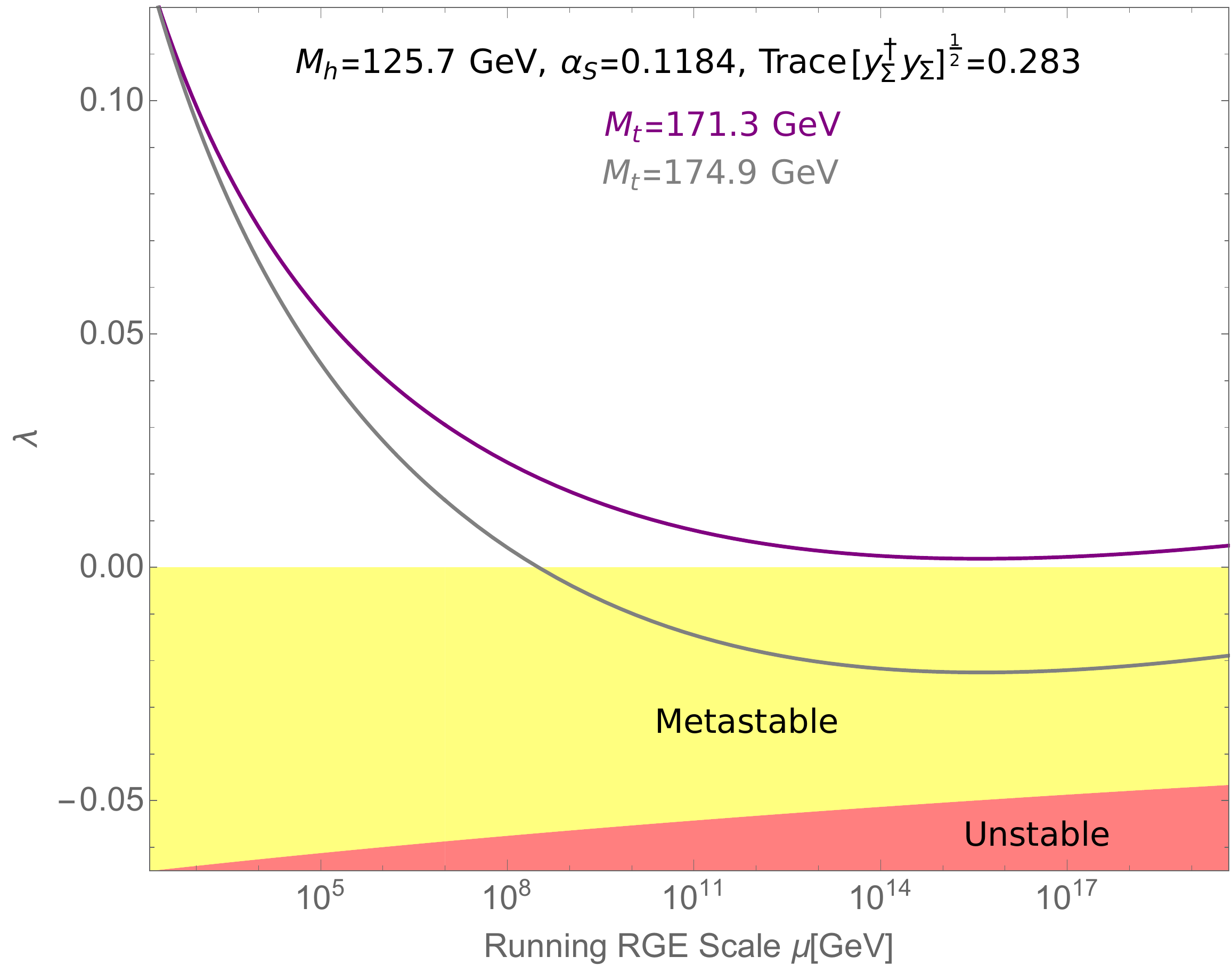}
         \end{subfigure}
       \begin{subfigure}[b]{0.469\textwidth}
        \includegraphics[width=\textwidth]{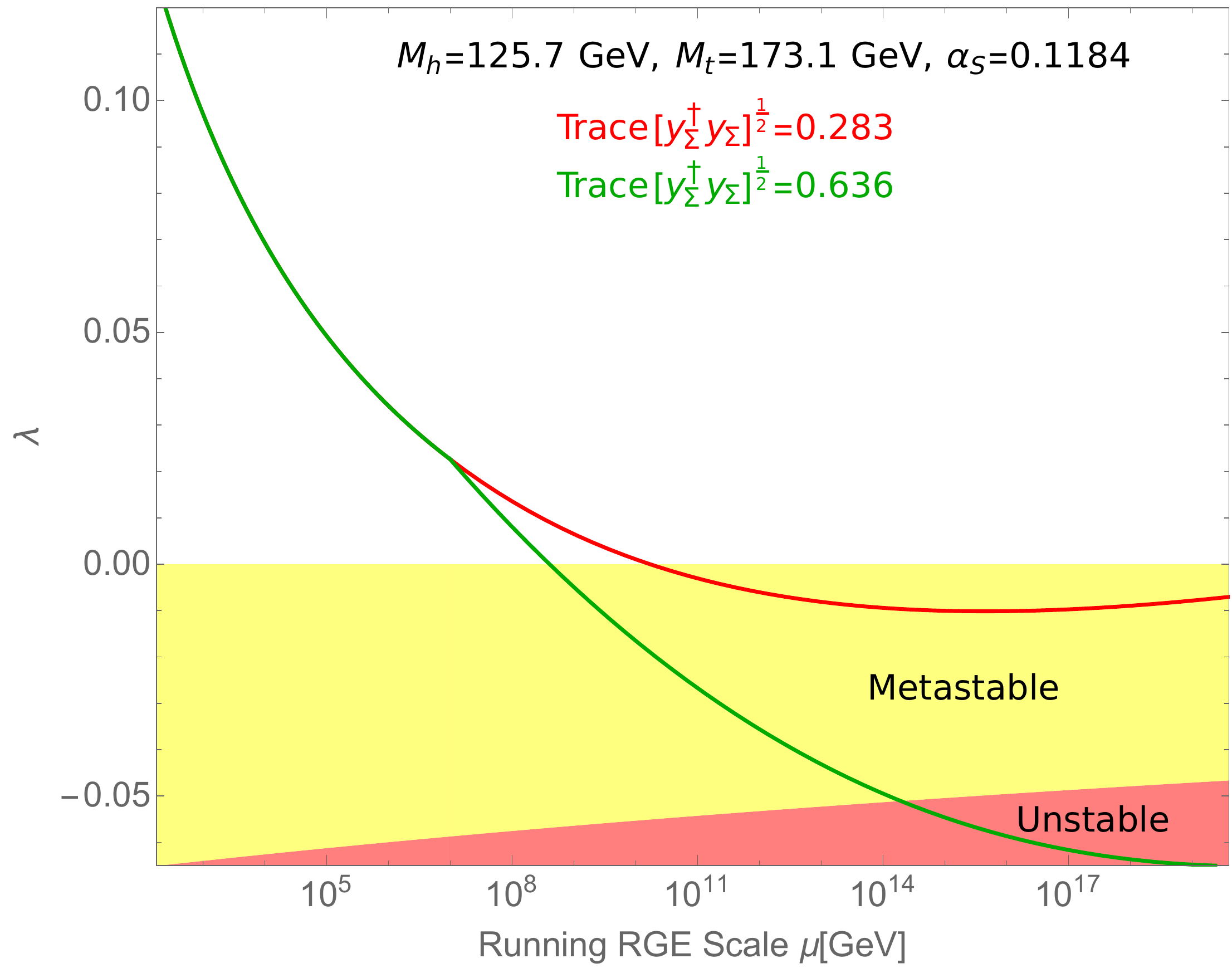}
                    \end{subfigure}
  \caption{RG evolution of the Higgs quartic coupling . The figure in the left side shows the running of $\lambda$ for different values of $M_t$ with fixed $M_\Sigma$  and Tr$[Y_{\Sigma}^\dag Y_\Sigma]^{\frac{1}{2}}$ whereas the figure in the right side shows the running of $\lambda$ for different values of Tr$[Y_{\Sigma}^\dag Y_\Sigma]^{\frac{1}{2}}$ with $M_\Sigma$  and $M_t$ fixed. For both the plots, we have taken $M_{\Sigma 1}=M_{\Sigma 2}=M_\Sigma =10^7$ GeV.}\label{RGE}
\end{figure}

 In Fig.\ref{RGE}, we show the running of the Higgs quartic coupling  for four different sets of benchmark points for the minimal type-III seesaw model.
In the first figure, the purple and gray lines correspond to $M_t=171.3$ and $174.9$ GeV respectively with the value of Tr$[Y_{\Sigma}^\dag Y_\Sigma]^{\frac{1}{2}}$ fixed as $0.283$ and $M_\Sigma =10^7$ GeV. For the first case, we can see that the Higgs quartic coupling $\lambda$ remains positive up to the Planck scale, i.e., the EW vacuum is absolutely stable up to the $\mpl$. For  $M_t=174.9$ GeV, we can see that $\lambda\sim \lambda_{eff}$ becomes negative at the energy scale $\sim 10^{9}$ GeV, the so called instability scale $\Lambda_I$, and remains negative upto $\mpl$.
However, we find that the 
beta function of the Higgs quartic coupling $\beta_{\lambda} (\equiv dV(h)/dh)$ becomes zero around the energy scale $\sim 10^{17}$ GeV, which implies that there is an extra deeper minima at that scale and we have checked that the EW vacuum corresponding to this point is metastable. Similarly in the second figure, we have given the running of the quartic coupling for two different values of Tr$[Y_{\Sigma}^\dag Y_\Sigma]^{\frac{1}{2}}$ with fixed $M_t$ and $M_\Sigma$. We notice that as the value of the Tr$[Y_{\Sigma}^\dag Y_\Sigma]$ is increased from 0.283 to 0.636, the EW vacuum shifts from the metastable to the unstable region. In this way, the conditions of stability and metastability can put constraints on the allowed values of Tr$[Y_{\Sigma}^\dag Y_\Sigma]^{\frac{1}{2}}$.

   \begin{center}
\begin{figure}[tbh]
\includegraphics[scale=0.4]{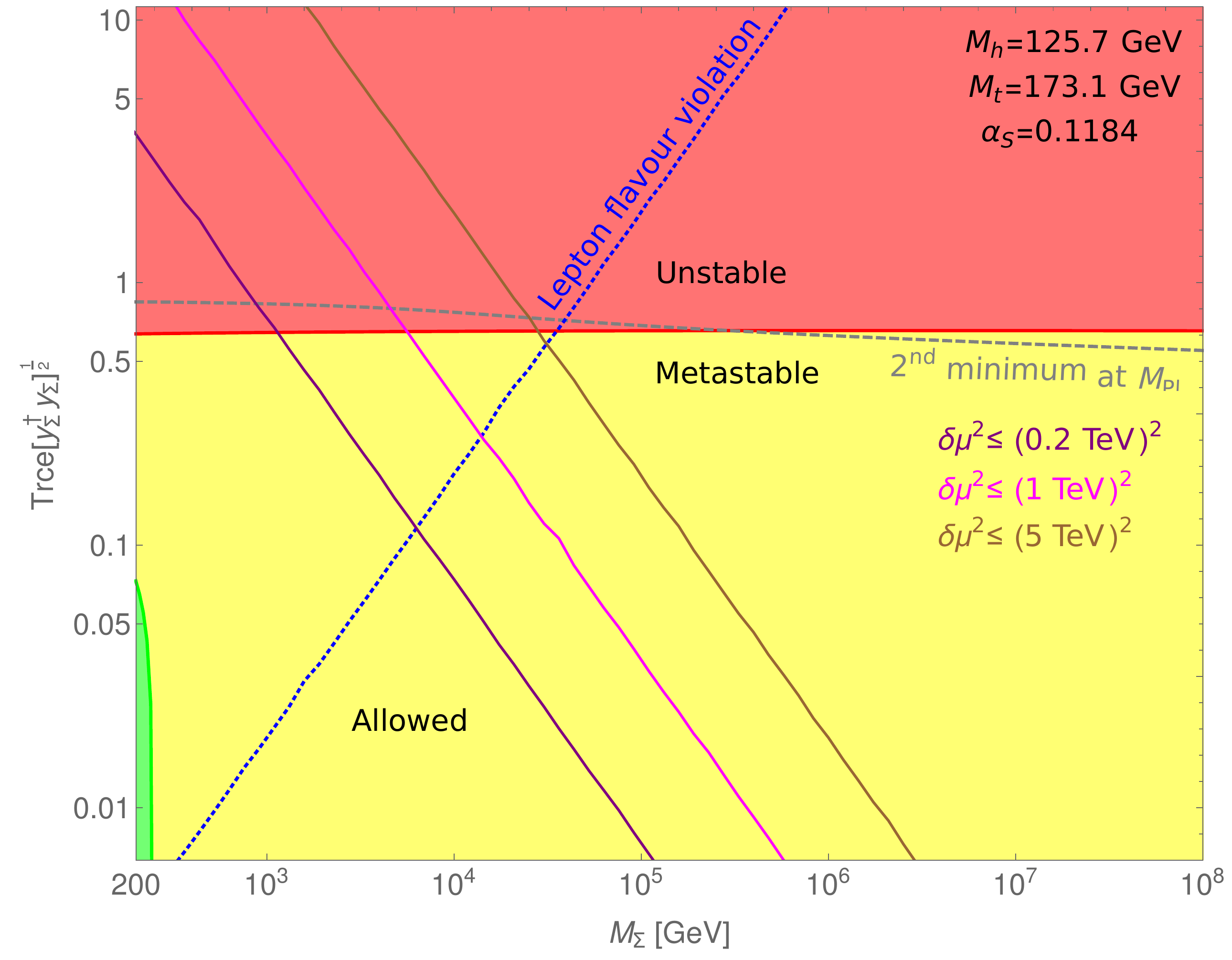}
 \caption{ The phase diagram in the $ \textrm{Tr}\, [Y_\Sigma^\dag Y_\Sigma]^{\frac{1}{2}} \,-\,M_\Sigma$ plane for NH. Here, we have used the central values of $M_t$, $M_h$ and $\alpha_s$. The color coding of the lines (blue, purple, magenta and brown) are the same as in figure \ref{3}. The horizontal red solid line separates the unstable and the metastable regions of the EW vacuum.}
    \label{flavorcentral}
\end{figure}
 \end{center} 
 \subsection{Phase diagram of Vacuum stability}
 
  \begin{figure}[h!]
    \begin{subfigure}[b]{0.495\textwidth}
        \includegraphics[width=\textwidth]{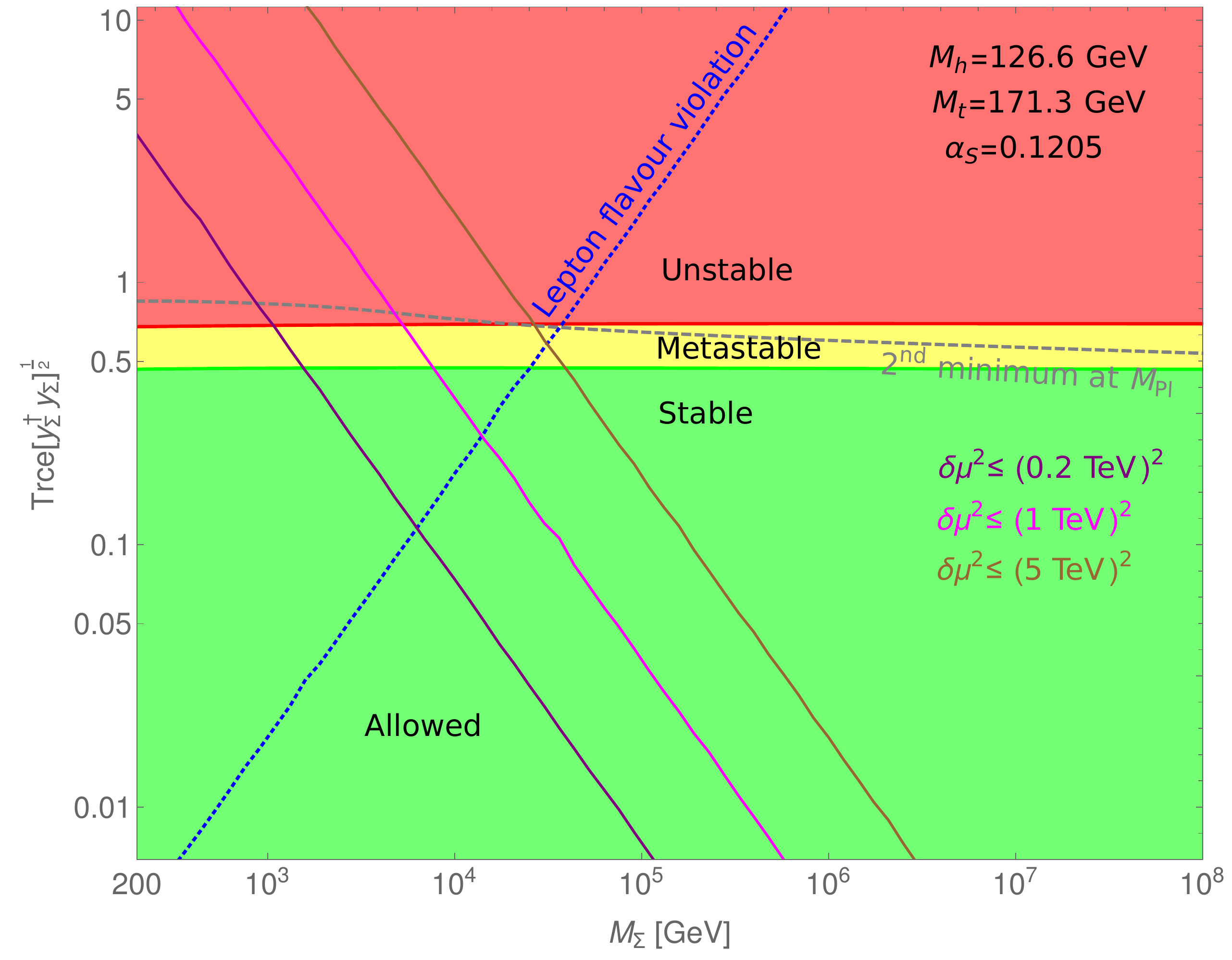}
         \end{subfigure}
       \begin{subfigure}[b]{0.495\textwidth}
        \includegraphics[width=\textwidth]{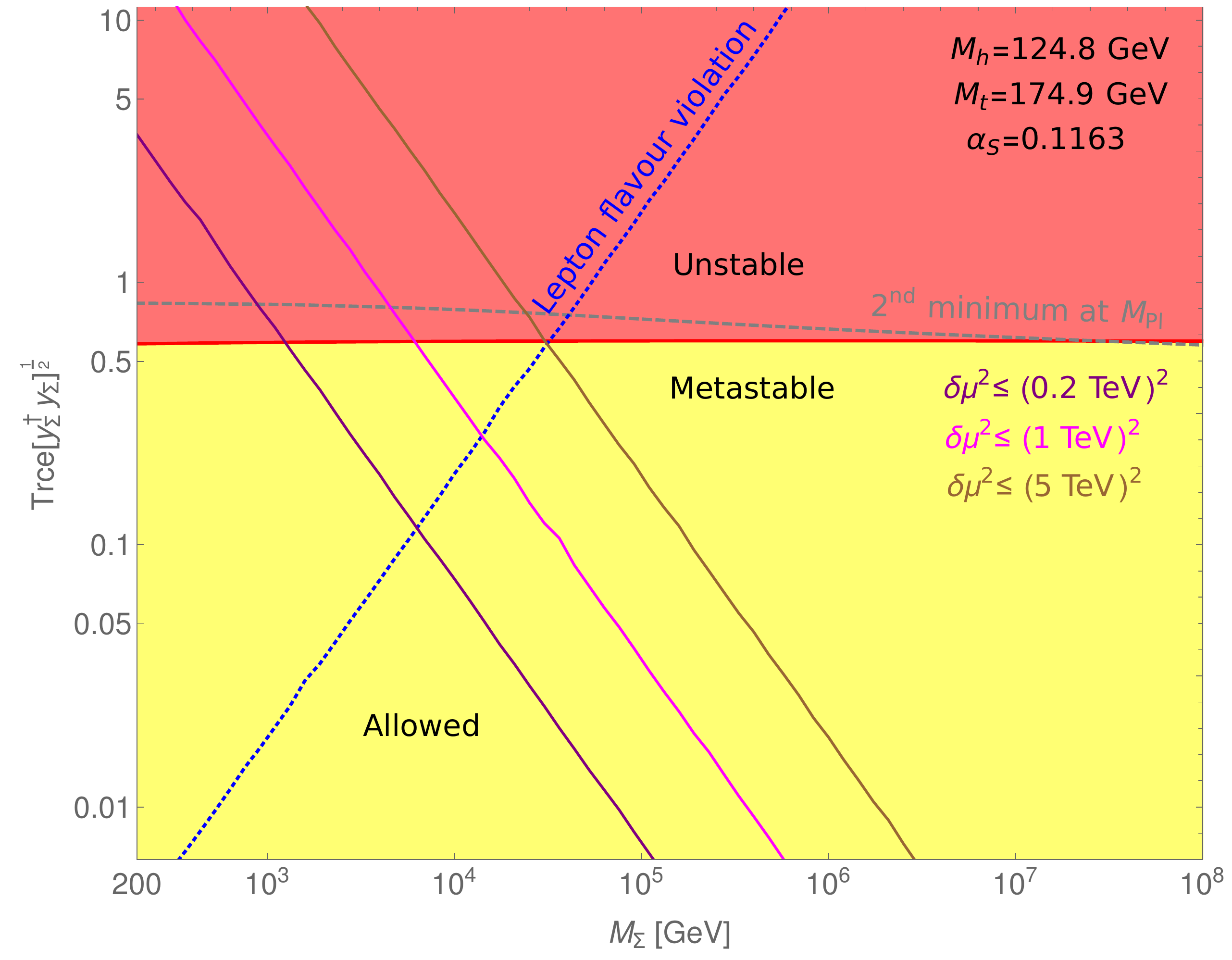}
                    \end{subfigure}
 \caption{ The phase diagram in the $ \textrm{Tr}\, [Y_\Sigma^\dag Y_\Sigma]^{\frac{1}{2}} \,-\,M_\Sigma$ plane for NH. The figure in the left (right) side gives the most liberal (stringent) bound from vacuum stability with minimum (maximum) value of $M_t$ and maximum (minimum) values of $M_h$ and $\alpha_s$. The color coding of the lines (blue, purple, magenta and brown) are the same as in figure \ref{3}. The horizontal red solid line separates the unstable and the metastable regions of the EW vacuum.}
    \label{flavormeta}
\end{figure}
 
As we have already discussed, the present central values of the SM parameters imply that an extra deeper minima exists near the Planck scale. Hence, there is a possibility that the EW vacuum might tunnel into that true (deeper) vacuum. In the type-III seesaw model, depending upon the  new physics parameter space, the stability of the EW vacuum is modified compared to that in the SM and there are two effects contributing to this. The first one is the negative contribution to the running of $\lambda$ as well as to the effective Higgs potential due to the triplet fermion Yukawa coupling (see the Eqns.~\ref{labmbdaeffnu} and \ref{eq:1betaHiggs}). The second one is through the modified RGE for the SU(2) gauge coupling, $g_2$ (Eqn.\ref{eq:1betag2}), which in turn gives a positive contribution to the running of $\lambda$. These effects have also been discussed in reference (\cite{Lindner:2015qva}).

 In Fig.\ref{flavorcentral}, we have given the phase diagram in the $ \textrm{Tr}\, [Y_\Sigma^\dag Y_\Sigma]^{\frac{1}{2}} \,-\,M_\Sigma$ plane for the central values of the SM parameters, $M_t = 173.1$, $M_{h} =  125.7$ and $\alpha_s=0.1184$. Here, the horizontal red solid line separating the unstable region (red) and the metastable (yellow) region is obtained when $\beta_\lambda(\mu) \, = \, 0$ along with $\lambda(\mu) \, = \, \lambda_{min}(\Lambda_B)$. From this plot, we can see that the parameter space with $\textrm{Tr}\,[ Y_\Sigma^\dag Y_\Sigma]^{\frac{1}{2}} \, \gtrsim \, 0.64 $ with the heavy fermion mass scale $200-10^{8}$ GeV are excluded by instability of the EW vacuum. The gray dashed line corresponds to the points for which the beta function of the quartic coupling $\lambda$ is zero at the Planck scale, i.e., the second minima is situated at that scale. Also, we can see a very small green region for lower values of masses and couplings for which the EW vacuum is absolutely stable. However, this region is disfavored from the LFV constraints as shown by the blue dotted line. The region to the right of this line is allowed by the current bounds from LFV as given in eqn.(\ref{lfvbound}). We have also given the bounds from naturalness in these figures
as shown by the slanted solid lines corresponding to three different values of $\delta\mu^2$. Thus, one can see that the area that are allowed both by naturalness as well as LFV falls in the stability/metastability region.
 
  In Fig.\ref{flavormeta}, we have again plotted the phase diagram in the $ \textrm{Tr}\, [Y_\Sigma^\dag Y_\Sigma]^{\frac{1}{2}} \,-\,M_\Sigma$ plane for NH, but with different values of the SM parameters. The figure in the left (right) side gives the most liberal (stringent) bound from vacuum stability with minimum (maximum) value of $M_t$ and maximum (minimum) values of $M_h$ and $\alpha_s$ from their allowed $3\sigma$ ranges. Clearly, with the smallest value of $M_t$ and the largest values of $M_h$ and $\alpha_s$, the stability region increases as is shown by the green region in the figure in the left-hand side. On the other hand, in the right panel with the highest value of $M_t$ and lowest values of $M_h$ and $\alpha_s$, no region of stability is found. In this case, the parameter space with  $ \textrm{Tr}\, [Y_\Sigma^\dag Y_\Sigma]^{\frac{1}{2}} > $  0.68 (0.58) is disfavored from the instability condition in the left(right) panels.

  \begin{center}
\begin{figure}[h!]
\includegraphics[scale=0.4]{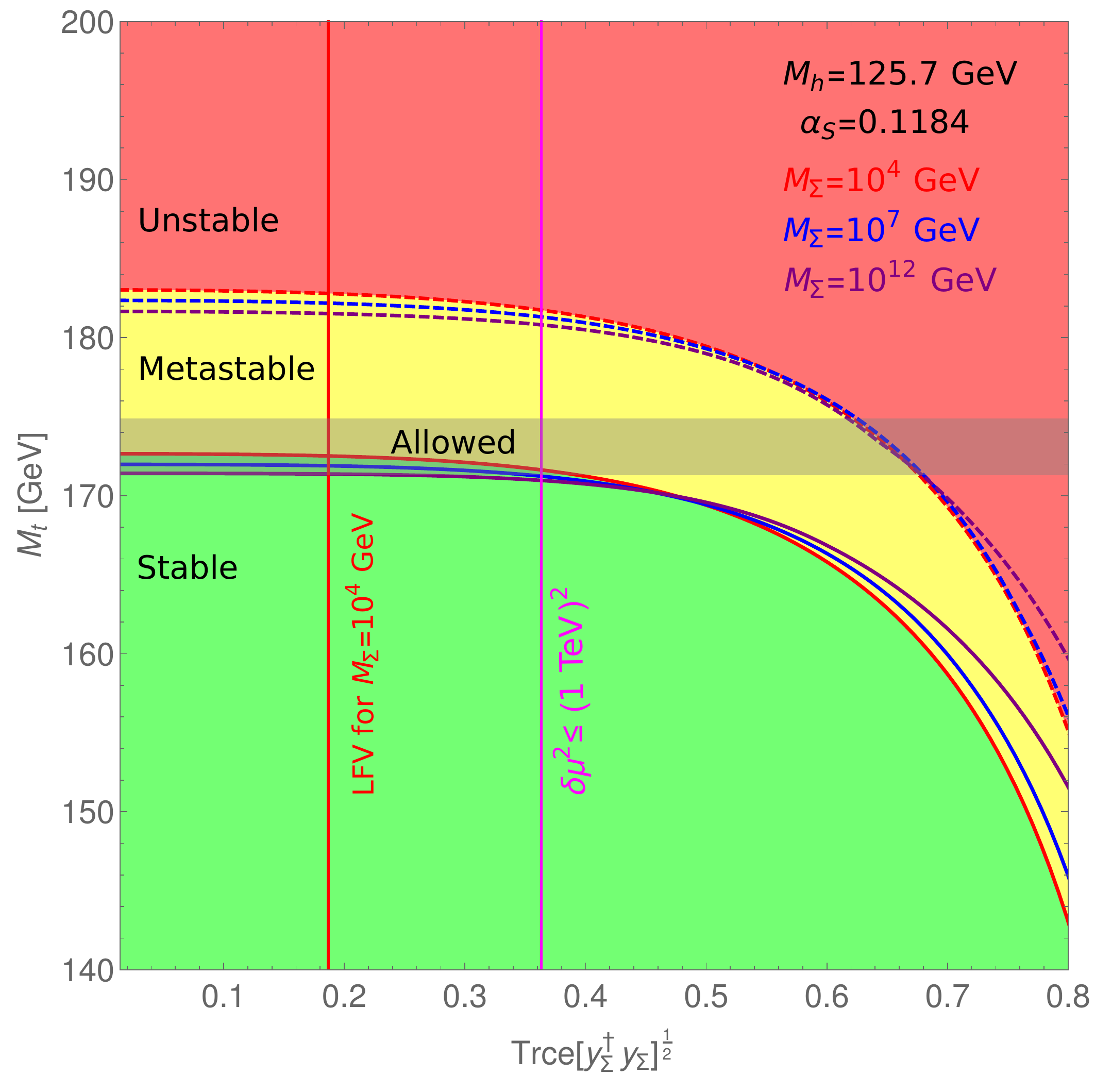}
 \caption{ The phase diagram in the $M_t- \textrm{Tr}\, [Y_\Sigma^\dag Y_\Sigma]^{\frac{1}{2}} $ plane for NH for the central values of $M_h$ and $\alpha_s$. The dashed lines separate the metastable and the unstable regions whereas the solid lines separate the stable and the metastable regions. The three colors are for for three different values of $M_\Sigma$. The two vertical lines give the LFV and naturalness bounds for $M_\Sigma = 10^4$ GeV and the region in the left of the LFV line (red) is allowed by both.}
 
    \label{MtTrYnuYnu}
\end{figure}
 \end{center} 
 Fig.\ref{MtTrYnuYnu} gives the phase diagram in the $M_t- \textrm{Tr}\, [Y_\Sigma^\dag Y_\Sigma]^{\frac{1}{2}} $ plane for NH with the central values of $M_h$ and $\alpha_s$. The dashed lines separate the metastable and the unstable regions whereas the solid lines separate the stable and the metastable regions. The red, blue and purple colored lines correspond to the representative values of $M_\Sigma$ as $10^4$, $10^{7}$ and $10^{12}$ GeV respectively. The two vertical lines give the LFV and the naturalness ($\delta\mu^2 < 1$ $\textrm{TeV}^2$) bounds for $M_\Sigma = 10^4$ GeV and the allowed region is to the left of the red vertical line. The horizontal shaded gray region denote the $3\sigma$ allowed range of $M_t$. It is seen that in this region, the vacuum is metastable for lower values of $ \textrm{Tr}\, [Y_\Sigma^\dag Y_\Sigma]^{\frac{1}{2}} $, while for higher values, the vacuum is unstable. Once we consider the bounds from LFV, $ \textrm{Tr}\, [Y_\Sigma^\dag Y_\Sigma]^{\frac{1}{2}} $ is less than 0.18 and the vacuum is in the metastable region.

\begin{figure}[tbh]
    \begin{subfigure}[b]{0.47\textwidth}
        \includegraphics[width=\textwidth]{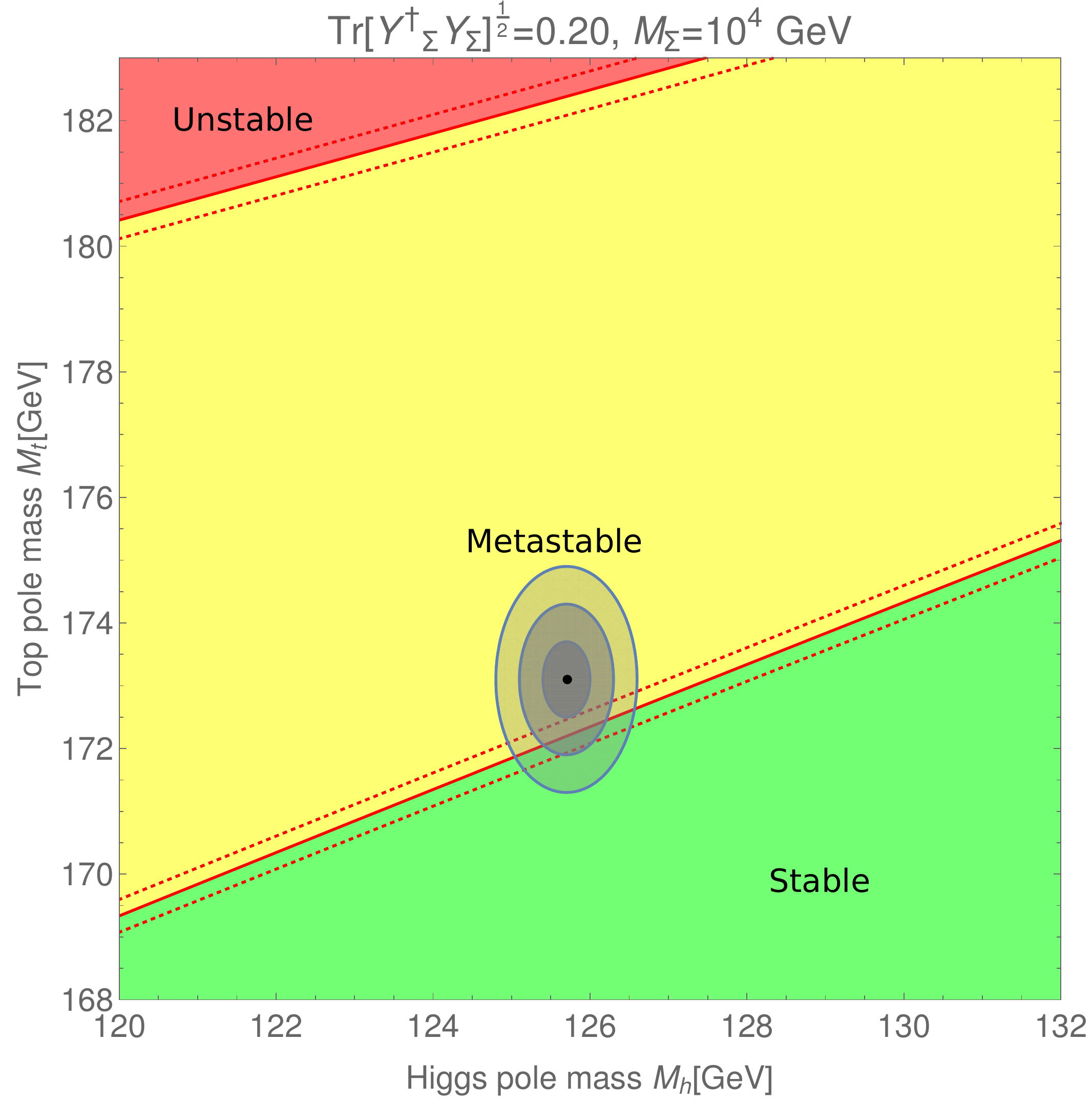}
        \caption{ \centering  $ \textrm{Tr}\, [Y_\Sigma^\dag Y_\Sigma]^{\frac{1}{2}} = 0.2$  }\label{8a}
         \end{subfigure}
       \begin{subfigure}[b]{0.47\textwidth}
        \includegraphics[width=\textwidth]{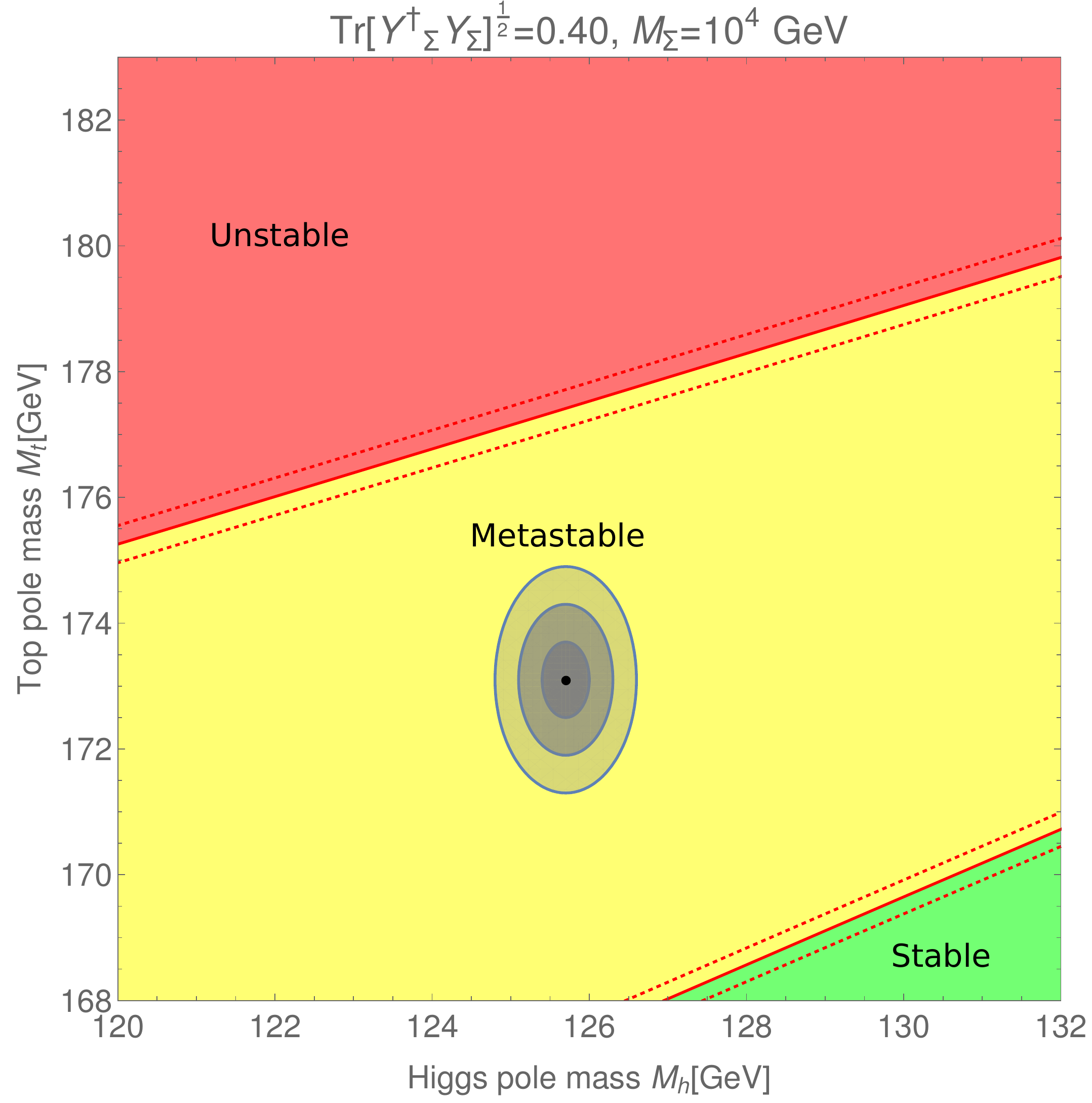}
        \caption{\centering $ \textrm{Tr}\, [Y_\Sigma^\dag Y_\Sigma]^{\frac{1}{2}}= 0.4$  }\label{8b}
                    \end{subfigure}
  \caption{The phase diagram in the $M_t-M_h$ plane for two different values of $ \textrm{Tr}\, [Y_\Sigma^\dag Y_\Sigma]^{\frac{1}{2}}$ and $M_\Sigma = 10^4$ GeV. The ellipses correspond to the allowed values of $M_t$ and $M_h$ at $1\sigma$, $2\sigma$ and $3\sigma$.}\label{mtmh}
\end{figure}

In Fig.\ref{mtmh}, we have shown the phase diagram in the $M_t-M_h$ plane for  $M_\Sigma = 10^4$ GeV. The red dashed lines correspond to the $3\sigma$ variation in $\alpha_s$. The figures in the left- and right-hand sides correspond to  $ \textrm{Tr}\, [Y_\Sigma^\dag Y_\Sigma]^{\frac{1}{2}} = $ 0.20 and 0.40 respectively. The ellipses correspond to the allowed values of $M_t$ and $M_h$ at $1\sigma$, $2\sigma$ and $3\sigma$. From this figure, we can clearly see that higher values of $M_t$ and $Y_\Sigma$ affect the stability of the EW vacuum negatively whereas higher value of $M_h$ has a positive effect on the stability. For $ \textrm{Tr}\, [Y_\Sigma^\dag Y_\Sigma]^{\frac{1}{2}} = $ 0.20, some areas of the parameter space fall in the stable region when $M_t$ and $M_h$ are taken in the $3\sigma$ ranges, whereas for  $ \textrm{Tr}\, [Y_\Sigma^\dag Y_\Sigma]^{\frac{1}{2}} = $ 0.40, all the allowed parameter space is in the metastable region.   

  \begin{figure}[h!]
    \begin{subfigure}[b]{0.47\textwidth}
        \includegraphics[width=\textwidth]{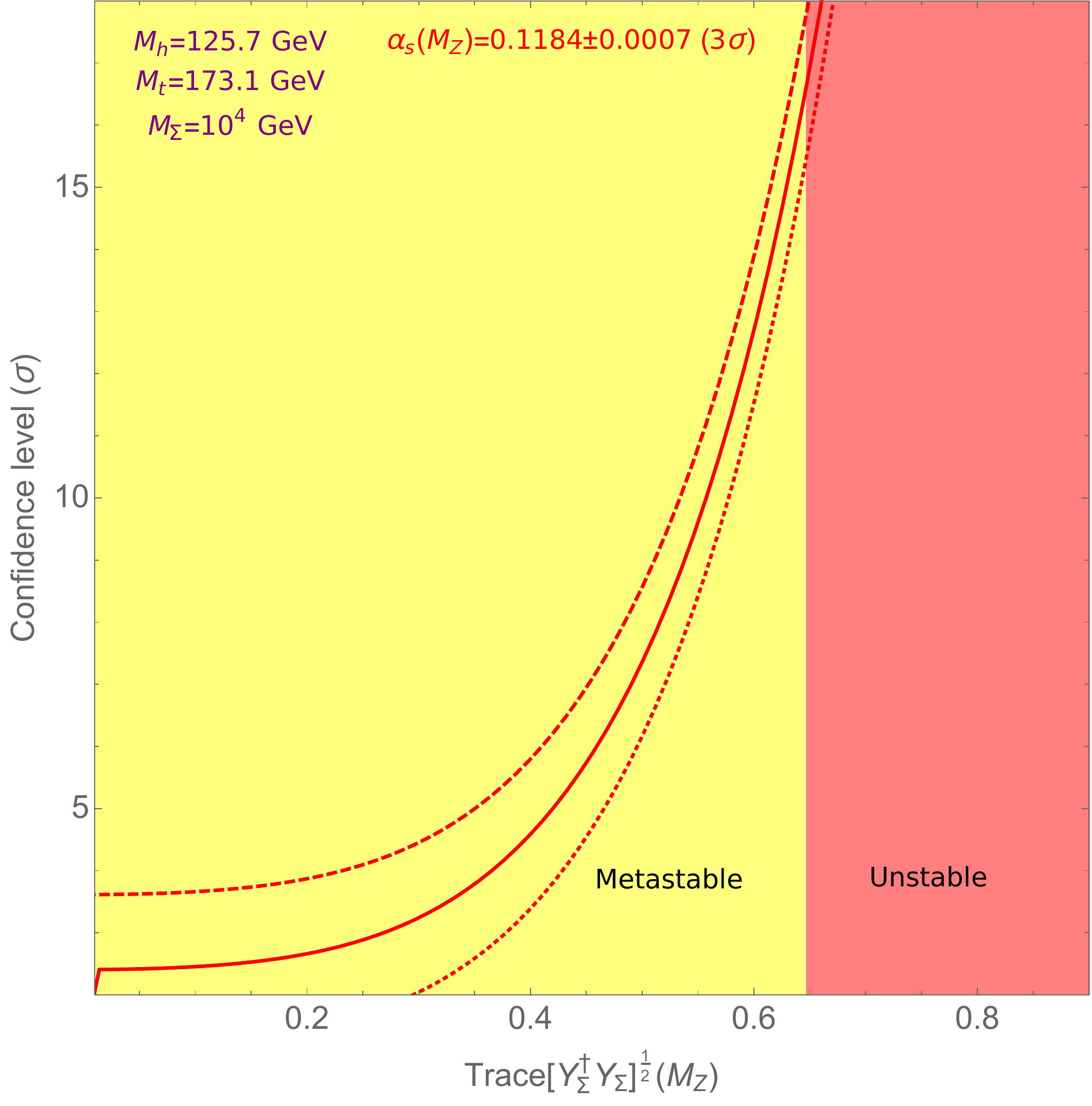}
        \caption{ \centering  $M_\Sigma = 10^4$ GeV   }\label{3a}
         \end{subfigure}
       \begin{subfigure}[b]{0.47\textwidth}
        \includegraphics[width=\textwidth]{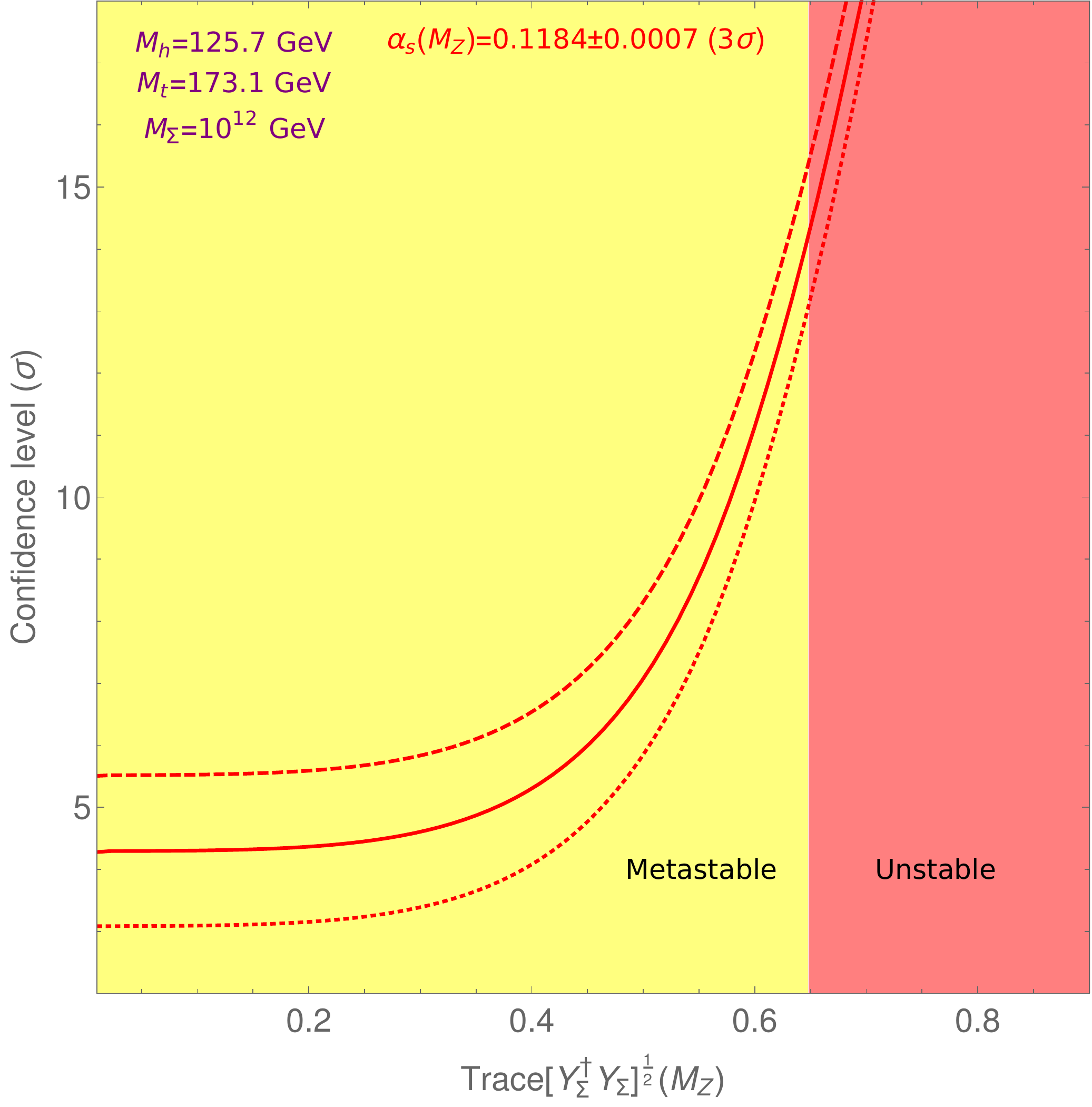}
        \caption{\centering $M_\Sigma = 10^{12}$ GeV  }\label{3b}
                    \end{subfigure}
  \caption{Dependence of confidence level at which the EW vacuum stability is excluded/allowed on $ \textrm{Tr}\, [Y_\Sigma^\dag Y_\Sigma]^{\frac{1}{2}}$ for different values of $\alpha_s$ and $M_\Sigma$.}\label{CDISM}
\end{figure}

It is also important to look at the change in the confidence level at which the (meta)stability is excluded or allowed \cite{Khan:2014kba, Khan:2015ipa, Khan:2016sxm} in the context of the minimal type-III seesaw model. The confidence level plot(s) will provide a quantitative measurement of the (meta)stability for the new physics parameter space.
In Fig.\ref{CDISM}, we show how the confidence level at which EW vacuum is allowed(excluded) from the metastability(instability) depends on new Yukawa couplings of the heavy fermions for the type-III seesaw model for different values of $M_\Sigma$ and $\alpha_s$. To plot these, we have considered the variation of $M_t$ (from 160 to 180 GeV) and $M_h$ (from 120 to 132 GeV) in the $M_t-M_h$ plane for fixed values of $\alpha_s$. We draw the metastability line and an ellipse to which the metastability line is the tangent and the point corresponding to the central values of $M_t$ and $M_h$ ($M_t= 173.1$ GeV, $M_h = 125.7$ GeV) as the center (See Fig.\ref{mtmh} for instance). Then we calculate the confidence level as,
$\textrm{Confidence level} = \frac{a\textrm{ of the ellipse}}{1 \sigma \textrm{ error of } M_t}  =  \frac{b\textrm{ of the ellipse}}{1 \sigma \textrm{ error of } M_h}$, where $a$ and $b$ are the lengths of the major and minor axes of the ellipse. Fig.\ref{3a} and Fig.\ref{3b} are plotted with the triplet masses as  $M_\Sigma=10^{4}$ GeV and $10^{12}$ GeV respectively. In both cases the EW vacuum is  metastable for smaller values of the new Yukawa coupling. We can see that the confidence level at which the EW  vacuum
is metastable (yellow region) increases with the increase of $ \textrm{Tr}\, [Y_\Sigma^\dag Y_\Sigma]^{\frac{1}{2}}$. Also, one can see that the confidence level at which the EW vacuum is metastable increases with the increase in the mass of the fermion triplets. We can also see the effect of $\alpha_s$ on the confidence level. The dashed, solid and dotted red lines correspond to the values of $\alpha_s$ as 0.1177, 0.1184 and 0.1191 respectively. Clearly, the confidence level at which the EW vacuum is metastable decreases with the increase in $\alpha_s$. This is because, $\alpha_s$ has a positive effect on the stability of the EW vacuum and the increase in $\alpha_s$ increases the confidence level at which the vacuum is stable and thereby decreasing the confidence level at which it is unstable. The EW vacuum becomes metastable for $ \textrm{Tr}\, [Y_\Sigma^\dag Y_\Sigma]^{\frac{1}{2}} = 0.646 \pm 0.008 $ and $ \textrm{Tr}\, [Y_\Sigma^\dag Y_\Sigma]^{\frac{1}{2}} = 0.648 \pm 0.011 $ corresponding to $\alpha_s = 0.1184 \pm 0.0007 $ for $M_\Sigma = 10^4 \textrm{ and } 10^{12}$ GeV respectively. The demarcations between the stable and the metastable regions in the plots are only for the central values of $\alpha_s$

\section{Summary}

In this paper we have analyzed the implications of naturalness and the stability of the electroweak vacuum in the context of the minimal type-III seesaw model. We have also studied the constraints from lepton flavor violating decays. We have found that the lighter masses of the fermionic triplets, $M_\Sigma \simeq 400$ GeV are disallowed for all values of $Y_\Sigma$ by the constraints from the $\mu \rightarrow e$ conversion in the nucleus. At the same time, the heavier triplet masses are disfavored by naturalness. For instance, if we demand the correction to the Higgs mass to be less than 200 GeV, it will put an upper bound of $\sim 10^5$ GeV on the masses of the triplets. Also, the maximum value of $\textrm{Tr}[Y_\Sigma^\dag Y_\Sigma]^{\frac{1}{2}}$ that is allowed is 0.1, corresponding to $M_\Sigma \sim 10^4$ GeV.  Another important result is that in the parameter space which is allowed by both the LFV as well as naturalness constraints, the EW vacuum is stable/metastable depending on the values of $\textrm{Tr}[Y_\Sigma^\dag Y_\Sigma]^{\frac{1}{2}}$ and the standard model parameters used. Hence, one does not really have to worry about the instability of the vacuum in this model. The major part of the allowed parameter space lies in a region that could be tested in the future collider experiments.

\section*{Acknowledgement}

The work of Najimuddin Khan is supported by the Department of Science and Technology, Government of INDIA under the SERB-Grant PDF/2017/00372.

 \appendix
\section{Renormalization Group Equations}

The beta functions for the various couplings are defined as,
\bea
\beta_{\chi_{i}}=\frac{\partial \chi_{i}}{\partial \ln \mu} =   \frac{1}{16 \pi^2}~\beta_{\chi_{i}}^{(1)}  +  \frac{1}{(16 \pi^2)^2}~\beta_{\chi_{i}}^{(2)}\, .\nn 
\eea
For the running scale $\mu < M_\Sigma$,
\allowdisplaybreaks \be
\beta_{\chi_{i}} = \beta_{\chi_{i}}^{SM} ~~~{\rm ,} ~~~\beta_{g_2}^{(1)} = -\frac{19}{6}g_2^3 ~~~{\rm and}~~~ \beta_{Y_\Sigma} =0,\nn
\ee

and for $\mu>M_\Sigma$, the one-loop RGEs for $\lambda$, $y_t$, $g_2$ and $Y_\Sigma$ are as given below.

{\allowdisplaybreaks  
\begin{align}
\beta_{\lambda} & =  
\frac{3}{8} g_{1}^{4} +\frac{3}{4} g_{1}^{2} g_{2}^{2} +\frac{9}{8} g_{2}^{4} -3 g_{1}^{2} \lambda -9 g_{2}^{2} \lambda +24 \lambda^{2} + 12 \lambda y_t^2 -6 y_t^4 \nonumber \\ 
 & +12 \lambda \mbox{Tr}\Big({Y_\Sigma  Y_{\Sigma}^{\dagger}}\Big) -10 \mbox{Tr}\Big({Y_\Sigma  Y_{\Sigma}^{\dagger}  Y_\Sigma  Y_{\Sigma}^{\dagger}}\Big)\label{eq:1betaHiggs}\\
 \nn\\
 \beta_{y_t} & =  
y_t \Big( \frac{9}{2} yt^2  -8 g_{3}^{2}  -\frac{17}{12} g_{1}^{2}  -\frac{9}{4} g_{2}^{2} + 3 \mbox{Tr}\Big({Y_\Sigma  Y_{\Sigma}^{\dagger}}\Big)   \Big)\label{eq:1betaYuk}\\
\nn\\
 \beta_{g_2} & =  
-\frac{1}{2} g_2^3\label{eq:1betag2}\\
\nn\\
\beta_{Y_\Sigma} & =  
 Y_\Sigma \Big(\frac{5}{2} {Y_\Sigma  Y_{\Sigma}^{\dagger} } +  3 y_t^2  -\frac{33}{4} g_{2}^{2}  -\frac{3}{4} g_{1}^{2} + 3 \mbox{Tr}\Big({Y_\Sigma  Y_{\Sigma}^{\dagger}}\Big) \Big) \label{eq:1betaBSM}\\ 
\end{align}}

Two-loop RGEs used in this work have been generated using {\tt SARAH}~\cite{Staub:2013tta}. In our work, we have taken only the top-quark contributions. The other SM-Yukawa couplings are comparatively smaller and their inclusion does not alter our result.

\bibliographystyle{utphys}
\bibliography{tevportalnew}

\end{document}